\documentclass{article}
\usepackage{arxiv_adj}
\usepackage[figurename=Fig.]{caption}
\usepackage{hyperref}
\hypersetup{colorlinks=true,linkcolor=blue,citecolor=blue,filecolor=magenta,urlcolor=cyan}
\usepackage{amsmath,latexsym,amssymb,graphicx,dsfont,amsthm,amsfonts}
\usepackage[linesnumbered,ruled,vlined]{algorithm2e}
\SetKwInput{KwInput}{Input}        
\SetKwInput{KwOutput}{Output} 
\newcommand{\el}{\texttt{l}}
\newcommand{\N}{\mathbb{N}}

\begin{document}


\title{Robustness and efficiency of\\leaderless probabilistic consensus protocols\\within Byzantine infrastructures}


\author{Angelo Capossele$^*$, Sebastian Mueller$^+$, Andreas Penzkofer$^*$ \\
\small{$^*$ IOTA Foundation, 10405 Berlin, Germany} \\
\small{$^+$ Aix-Marseille Universit\'e, 
CNRS,  Centrale Marseille, 
I2M, UMR 7373, 13453 Marseille, France
}}

\maketitle

\begin{abstract}
This paper investigates leaderless binary majority consensus protocols with low computational complexity in noisy Byzantine infrastructures. Using computer simulations, we show that explicit randomization of the consensus protocol can significantly increase the robustness towards faulty and malicious nodes. We identify the optimal amount of randomness for various Byzantine attack strategies on different kinds of network topologies. 
\end{abstract}




\section{Introduction}

\subsection{Context}

Distributed consensus algorithms allow networked systems to agree on a required state or opinion in situations where centralized decision making is difficult or even impossible. As distributed computing is inherently unreliable, it is necessary to reach consensus in faulty or Byzantine infrastructure. The importance of this problem stems from its omnipresence and fault tolerance is one of the most fundamental aspects of distributed computing. Common examples include clock synchronization, resource allocation, task scheduling and replicated file systems, see \cite{BaDaMa:93}.

The consensus problem arises traditionally as well in technical systems for object tracking \cite{GeCi:96}, density classification \cite{MiHrCr:93}, sensor fusion \cite{OlSh:05}, robot localization \cite{SiKeBa:10}, and mission planning \cite{AlHo:06}. Furthermore, collective and distributed decision making is important in social dynamics \cite{CaFoLo:09, MoNeTa:13}. The consensus problem is also at the heart of distributed ledger technology, \cite{Zhetal:17}, since it allows to reach consensus on whether to add a transaction to a ledger or to agree on a timestamp of a transaction.

Every node in a distributed consensus system follows simple rules, using only local information and node-to-node communication. Algorithms performing such tasks should be robust toward different types of disturbances and Byzantine settings. A landmark study \cite{PeShLa:80} considers the robustness of consensus with faulty nodes. In particular, \cite{PeShLa:80} show that for synchronous systems a necessary condition for reaching consensus is that at most $1/3$ of the nodes are faulty. In the significant work \cite{FiLyPa:85} this result is strengthened in the sense that in asynchronous situations already one faulty node can prevent an agreement. 

\subsection{Distributed binary majority consensus}\label{sec:binaryconsensus}
This article focuses on a certain class of consensus, namely, binary majority consensus. Some basic algorithms in this protocol class are simple majority consensus, \cite{MoNeTa:13}, Gacs-Kurdyumov-Levin (GKL)\cite{GaKuLe:78}, and random neigbors majority, \cite{GoMaMaBe:15}. These protocols may achieve good performances in noiseless and undisturbed networks. However, their performances significantly decreases with noise \cite{GaKuLe:78, KaMo:07} or errors \cite{MoDiAm:04} and may completely fail in a Byzantine setting, see Section \hyperref[sec:SMCA]{\ref{sec:SMCA}}.
The sensitivity of these protocols to topological changes are, for example, studied in \cite{SaZa:93, KiBe:06, Beetal16}.
This weak robustness against faulty nodes may serve as an explanation of why simple majority voting was not thoroughly investigated until recently. 
In \cite{GoMaMaBe:15} several variants of the binary majority consensus were compared in networks with different types of disturbances. Moreover, \cite{GoMaMaBe:15} studied the potential capacity of randomization to improve the performance of the protocol.

\subsection{Novel contributions of fast probabilistic consensus}
While among the above disturbances the faulty node behavior appears to be the most severe, Byzantine or malicious nodes introduce the next challenge for consensus protocols and can be considered as the cutting edge in the area. Recently, \cite{Po:19} proposed a new leaderless binary majority consensus protocol, called the fast probabilistic consensus (FPC), and obtained theoretical results on the convergence and safety of the protocol. 
We propose a simplified version of the FPC that is easy to implement and contains nevertheless all ingredients that allow the protocol to be robust and of low communicational complexity in Byzantine infrastructures. A special feature of the protocol is that it makes use of a sequence of random numbers which are either provided by a trusted source~\cite{nist:2019} or generated by the nodes themselves using some decentralized random number generator protocol~\cite{popov-drng:2017, syta:2017, boneh:2018}. See Section \ref{sec:GeneratingRandNum} for more details.
 
Randomized algorithms are used in various fields and are known to offer efficient ways to reduce complexity and increase performances. In our context, a deterministic consensus protocol might fail in various situations, e.g.\ see \cite{GoMaMaBe:15} and Section \hyperref[sec:SMCA]{\ref{sec:SMCA}}. While such situations might be prevented by detection methods, these techniques are not always available or not yet developed for Byzantine infrastructure. Simple consensus protocols without randomization nor fault detection can rarely guarantee the eventual agreement of the nodes. However, noisy environment and random topology may lead to better performances, see \cite{SaShWo:91, As:02, MoDiAm:04}. 
In addition to these effects, the randomization in our protocol serves as ``fog of war'', \cite{vCl}, for potential malicious attackers, making it at best impossible for the attacker to control the honest nodes. However, let us note that the additional randomness may have some negative side effects: it can delay the termination of the protocol and lower the integrity rate, see Section \hyperref[sec:sim:Int]{\ref{sec:sim:Int}} for a more detailed discussion.

The FPC contains two additional new features. Firstly, it is adapted in the sense that nodes follow a local stopping rule on when to stop querying other nodes. Once all nodes stopped querying, the protocol terminates automatically.
Secondly, the protocol is asymmetric in the following sense. The two binary choices, $0$ and $1$, do not need to be of the same importance. For instance, in various applications, it may be appropriate to distinguish between the following two different kinds of errors. If initially, a majority of the nodes prefer $0$ but the eventual outcome of the protocol is $1$, and on the other hand if initially, a majority of the nodes prefer $1$ but the eventual outcome of the protocol is $0$. This distinction between the different kinds of errors is a standard approach in statistical test theory and in the evaluation of binary classifiers. 
The two different kinds of errors of a binary decision rule or classifier are measured differently in different fields. For example, in medicine, sensitivity and specificity are often used, in statistical test theory, one speaks of errors of type I and type II, and in computer science, precision and recall are preferred.

Finally, let us stress further important properties of the studied protocols. Nodes use only local information on the network, allowing the protocol to run on permissionless networks. Moreover, there is no central entity, nor elected leader that ``supervises'' the network and that decides whether consensus was achieved. This means that each node must decide when to stop using a local rule, i.e., using only the information locally available to it.

Byzantine infrastructures and attack strategies can be very diverse. We assume that adversarial nodes can exchange information freely between themselves and can agree on a common strategy. In fact, they all may be controlled by a single individual or entity.

\subsection{Key contributions}

In Section \ref{sec:protocols} we compare several leaderless binary majority consensus protocols, namely the simple majority consensus, the random majority consensus and the FPC. We show that the former two can be understood as a special case of the FPC. Furthermore, we propose a variation of FPC with reduced complexity by removing the randomness of the initial threshold and by omitting the cooling phase.
Since the theoretical bounds on the performances on FPC obtained in \cite{Po:19} are not optimal, we perform a thorough simulation study of FPC with various different parameter settings.

Our focus lies in the performances of the different consensus protocols in Byzantine infrastructures. For this purpose we propose in Section \ref{sec:faultyMaliciousNodes} explicit cautious and Berserk adversarial strategies. These strategies serve as a benchmark to compare the different protocols. The performances of the protocols are measured using integrity rate, agreement rate and termination rate. Furthermore, we subdivide the integrity rate into 0- and 1- integrity rate in order to take into account the asymmetric nature of the protocol, see Section \ref{sec:sim:Int}.

In \cite{Po:19}, it is assumed that every node holds a complete network view. However, in permissionless systems or in less reliable networks with inevitable churns (nodes join and leave) this assumption is not necessarily verified. We define in Section \ref{sec:networktopology} several graph topologies modeling the partial network view for the nodes and study their influence on the performances of the protocols. In particular, we show that FPC can perform well if nodes only have a partial network view.

We show that the random threshold of the FPC is necessary to allow the protocol to withstand a positive proportion of nodes that follow a Berserk strategy, see Fig.\ \ref{fig:eta-heatmap2}. Furthermore, we identify the optimal amount of randomness in the common random threshold in various situations, see Section \ref{sec:sim:TermAgree}.
We also note that the protocol allows some flexibility on this feature since it is not necessary to provide a common random threshold for every round, see Figs.\ \hyperref[fig:rand-TAI]{\ref{fig:rand-TAI}} and \hyperref[fig:rand-TAI]{\ref{fig:rand-rounds}}.

We show that the protocol exhibits good scaling behavior, with a message complexity of essentially $O(n)$ even in Byzantine infrastructure, see Figs. \hyperref[fig:N-TAI]{\ref{fig:N-TAI}} and \hyperref[fig:N-rounds]{\ref{fig:N-rounds}}.

A more detailed summary of the conclusions from the simulation analysis is given in Section \ref{sec:conclusions}.

\subsection{Simulation analysis}

In this section, we present an overview of the methodology for our simulation studies in Section \hyperref[sec:simulation]{\ref{sec:simulation}}. 

In order to assess the performance of the protocol we define several metrics, that evaluate the resilience against failures as well as the message complexity in Section \ref{sec:performanceMeasures}. A traditional way to measure the performance of the studied protocols is to consider the agreement rate, i.e.~the fraction of time the protocol reaches agreement among the non-faulty honest nodes. In order to obtain a more complete picture, we also study termination and integrity rates. We define termination rate as the rate at which the protocol reaches consensus between all nodes within a reasonable time. Furthermore, integrity rate is the rate at which the final opinion equals the initial majority opinion. 

In Section \hyperref[sec:sim:network-partitioning]{\ref{sec:sim:network-partitioning}} we analyze the performance of the protocol if nodes do not have a complete network view or if the network is not fully connected. To enable this we employ the Watts-Strogatz graph to model the network topology. This model allows to interpolate the topology between a worst-case scenario, in this case, a ring graph and the complete graph. 

Since a Byzantine environment is more severe than faulty nodes or message loss we focus our study on the former. In Section \hyperref[sec:maliciousNodes]{\ref{sec:maliciousNodes}}, three different types of adversarial strategies are introduced and applied to investigate the performance of the protocol. 

In Section \hyperref[sec:sim:Int]{\ref{sec:sim:Int}} we study the integrity of the protocol by applying the initial minority vote strategy, which aims to achieve an integrity failure. Due to the introduced asymmetry, for the integrity of 0- and 1-opinions, we subdivide the discussion on the integrity rate into whether integrity is maintained when the initial honest majority is 1 or 0. Under these conditions, we analyze the impact of the added asymmetry, the scalability of the protocol as well as up to which proportion of adversary nodes the protocol performs well. 

In Section \hyperref[sec:sim:TermAgree]{\ref{sec:sim:TermAgree}} we compare two strategies that aim for agreement and termination failure. In the inverse voting strategy, we assume the adversary may not send differing opinions to different nodes, while in the maximal variance strategy we allow this capability for the adversary. We call this capability cautious and Berserk, respectively. We compare the severity of these two different approaches, how the randomness of the protocol can overcome these attacks, and whether the protocol still performs well when the randomness is not continuously supplied.

\subsection{Further related work}
There is a wide range of classical work on (probabilistic) Byzantine consensus protocols \cite{AgTo:12, Be-0:03, Br:87, FeMi:89, FrMoRa:05, Ra:83}. A disadvantage of the approaches of these papers is, however, that they typically require that every node communicates with every other node in the network. This induces a communication complexity of $O(n^2)$ messages in each round. Moreover, in permissionless, unreliable or large networks, nodes may only be able to communicate with a subset of the network.

Most of the previous research focuses on failures within a network infrastructure, rather than on malicious agents. The work \cite{Lietal:18} defines a fast and scalable Byzantine fault-tolerance protocol. A leaderless Byzantine consensus is studied in 
\cite{Cretal:18} that provides a robust infrastructure where there is a failure in the leader of the consensus network. Another Byzantine fault tolerant method that does not require a leader node is Honey Badger \cite{Mietal:16}.

\section{Protocols}\label{sec:protocols}
In order to define the consensus protocols we introduce some notation.
We assume the network to have $n$ nodes indexed by $1,2,\ldots, n$. Every node $i$ has an opinion or state. We note $s_{i}(t)$ for the opinion of the node $i$ at time $t$. Opinions take values in $\{0,1\}$. Note that in some references, e.g. \cite{GoMaMaBe:15}, this is chosen to be $\{-1,1\}$.

At each time step each node chooses $k$ (random) neighbors $C_{i}$, queries their opinions and calculates 
\begin{equation*}
\eta_{i}(t+1)=\frac1{k_{i}(t)} \sum_{j\in C_{i}} s_{j}(t),
\end{equation*}
where $k_{i}(t)$ is the number of replies at time $t$.

Any binary decision, as in statistical test theory or binary classification, gives rise to two possible errors. In many applications, the two possible errors are asymmetric in the sense that one error is more severe than the other. For instance, in statistical test theory, one speaks of errors of type I type
II, while in computer science, precision and recall are preferred. Choosing the threshold of the first voting round $\tau \in (0.5,1]$ 
allows the protocol to include this kind of asymmetry.

\subsection{Simple majority consensus (SMC) }\label{sec:SMCA}
We set $C_{i}=\mathcal{N}_{i}$, i.e. all neighbors are queries. The first opinion update is defined by 
\begin{equation*}
s_{i}(1)=\left\{ \begin{array}{ll}
1, \mbox{ if } \eta_{i}(1) \geq \tau, \\
0, \mbox{ otherwise,}
\end{array}\right. 
\end{equation*}
and for $t\geq 1$:
\begin{equation*}
s_{i}(t+1)=\left\{ \begin{array}{ll}
1, \mbox{ if } \eta_{i}(t+1) > 0.5, \\
0, \mbox{ if } \eta_{i}(t+1) < 0.5, \\
s_{i}(t), \mbox{ otherwise.}
\end{array}\right. 
\end{equation*}

This kind of protocol is also known as ``cellular automata model'', see \cite{TaKiFu:96, WoOl:08, MaRo:08}, as majority dynamics, see e.g.\ \cite{MoNeTa:13, AbMo:15, Beetal16, GaZe:18} and in a more general setting as the Ising model, see \cite{Osetal:10}.

Without any additional security measures, this protocol is not only fragile against malicious attackers but can even produce agreement failures without disturbances, e.g.\, \cite{Beetal16}. A variation, called GKL, was proposed in \cite{GaKuLe:78} to avoid certain security features but stays extremely fragile in Byzantine infrastructures.

\subsection{Random neighbors majority consensus (RMC) } \label{sec:rmc}
The updated opinion  of a node is no longer computed using information from all its neighbors but using information from some randomly selected neighbors. A key feature is that neighbor selection is dynamic in the way that each node chooses different neighbors in each step $t$. This feature allows the RMC to be secure in various situation where the SMC would lead to an agreement or termination failure. At each time each node $i$ running uses the opinions from $C_{i}(t)$ random neighbors to update its own opinion: 
\begin{equation*}
s_{i}(1)=\left\{ \begin{array}{ll}
1, \mbox{ if } \eta_{i}(1) \geq \tau, \\
0, \mbox{ otherwise,}
\end{array}\right. 
\end{equation*}
and for $t\geq 1$:
\begin{equation*}
s_{i}(t+1)=\left\{ \begin{array}{ll}
1, \mbox{ if } \eta_{i}(t+1) > 0.5, \\
0, \mbox{ if } \eta_{i}(t+1) < 0.5, \\
s_{i}(t), \mbox{ otherwise.}
\end{array}\right. 
\end{equation*}

We assume that the $C_{i}(t)$ are independent for $i\leq n$ and $t\in\N$.
At each time $t$ for each $i$ the neighbors $C_{i}$ of a node $i$ are selected in a sample from a discrete uniform distribution on $\mathcal{N}_{i}$ without replacement. 

Let us note that in previous works, e.g. \cite{GoMaMaBe:15}, neighbors $C_{i}$ of a node $i$ were chosen using sampling with replacement and hence repetitions are possible. Our choice of using selection without replacement, facilitates the comparison with the following consensus protocol. Moreover, if the number of vertices is much higher than the number of queried neighbors, the difference between these two different choices is negligible. 
 
In \cite{GoMaMaBe:15} also the own state was taken into consideration and the protocol uses a variation for the calculation of the $\eta$'s:
\begin{equation*}
\widetilde \eta_{i}(t+1)=\frac1{k_{i}(t)+1} \left( s_{i}(t) +\sum_{j\in C_{i}} s_{j}(t)\right),
\end{equation*}
where $k_{i}(t)$ is the number of replies at time $t$. We note that in preliminary studies we did not detect any significant differences between these two versions for the values of $k$ under consideration.

\subsection{Fast probabilistic consensus (FPC)}
We consider a special version of the FPC introduced in \cite{Po:19} in choosing some parameters by default. Specifically, we remove the cooling phase of FPC and the randomness of the initial threshold $\tau$. The essential additional ingredient compared to RMC is that the threshold of $0.5$ for times $t\geq 2$ becomes random. Let $U_{t}, t=1, 2,\ldots$ be i.i.d.~random variables with law $\mathrm{Unif}( [\beta, 1-\beta])$ for some parameter $\beta \in [0,1/2]$. The update rules are now given by
\begin{equation*}
s_{i}(1)=\left\{ \begin{array}{ll}
1, \mbox{ if } \eta_{i}(1) \geq \tau, \\
0, \mbox{ otherwise,}
\end{array}\right. 
\end{equation*}
and for $t\geq 1$:
\begin{equation*}
s_{i}(t+1)=\left\{ \begin{array}{ll}
1, \mbox{ if } \eta_{i}(t+1) > U_{t}, \\
0, \mbox{ if } \eta_{i}(t+1) < U_{t}, \\
s_{i}(t), \mbox{ otherwise.}
\end{array}\right. 
\end{equation*}
Note that the RMC is a special case ($\beta=0.5$) of the FPC. We provide a pseudo-code of FPC in Algorithm \hyperref[alg:FPC]{\ref{alg:FPC}}.

\subsection{Termination of the consensus protocols}

\begin{algorithm}[t]
\DontPrintSemicolon
 
 \KwInput{opinionOld}
 \KwOutput{opinionNew}
 u := UNIF($\beta$,$1-\beta$)\;
  \While{ {\upshape(LENGTH(undecided) $>$ 0) AND (iteration $\leq$ maxIt)}}
   {
   		\For{ {\upshape node in undecided}}
		{c:= SAMPLE($\mathcal{N}\setminus$node, k, replace=FALSE) \;
		 eta:= MEAN(opinionOld[c])\; 
 \uIf{{\upshape(eta $>$ u)}}
    {       opinionNew[node]:= 1   
   }
   \uElseIf{{\upshape(eta $<$ u)}}
   {    opinionNew[node]:= 0  
  }
   \Else  {
  	opinionNew[node]:= opinionOld[node]}	
  \If{{\upshape(opinionOld[node]=opinionNew[node])}}
    {cnt++}
    \Else {cnt:= 0}
   
   \If{\upshape(cnt[node]=l)}
   {  	undecided:= undecided$\setminus$node	\;}
   }
   iteration++
}
\caption{FPC update for $n\geq 2$} \label{alg:FPC}
\end{algorithm}

We introduce local termination rules to reduce the communication complexity of the protocols. Every node keeps a counter variable \verb?cnt? that is incremented by $1$ if there is no change in its opinion and that is set to $0$ if there is a change of opinion. Once the counter reaches a certain threshold $\verb?l?$, i.e. $\verb?cnt?\geq \verb?l?$,  the node considers the current state as final. The node will therefore no longer send any queries but will still answer incoming queries. In case of no autonomous termination the algorithms is halted after $\verb?maxIt?$ iterations.

\section{Modelling assumptions}

\subsection{Network topology}\label{sec:networktopology}
We consider networks with $n$ nodes and where the nodes are enumerated from $1$ to $n$. Some node pairs are connected by a link/edge. For a given node $i$ we denote the set of all neighbors of $i$ by $\mathcal{N}_{i}$. 
We consider the following network topologies:
\begin{enumerate}
\item Complete graph: every node is connected to every other node
\item Regular ring lattice: In the regular ring lattice, introduced in \cite{GaKuLe:78}, every node is connected to an even number $d$ of other nodes in the following way. The $n$ nodes are numerated as $\{1,\ldots, n\}$ and each node connected to $d/2$ nodes on its ``left'' and $d/2$ nodes on its ``right'', i.e. there is an edge between $i$ and $j$ if and only if 
\begin{equation*}
0 < |i-j|~\mathrm{mod }~(n- k/2) \leq k/2.
\end{equation*}
We denote the proportion of the network a node is connected to by \begin{equation*}
\delta = d/N
\end{equation*}
\item Watts-Strogatz graphs: The Watts-Strogatz model, introduced in \cite{WaSt:98} is a random graph generation model that produces graphs with small-world properties, including short average path lengths and high clustering. A Watts-Strogatz graph with $n$ vertices and mean degree $d$ (even) is constructed as follows. Let $\gamma\in[0,1]$ be a model parameter and assume that $n \gg k \gg \ln n \gg 1$. The $nd/2$ undirected edges are now defined as follows:
\begin{enumerate}
\item Construct a regular ring lattice of degree $d$.
\item For every node $i=1,\ldots,n$ take every edge connecting $i$ to its $d/2$ rightmost neighbors and rewire it with probability $\gamma$. Rewiring here means that the edge is replaced by an edge between $i$ and a node chosen uniformly from the other nodes that are not yet neighbors of $i$. \end{enumerate}
The underlying ring lattice structure produces a locally clustered network, while the randomly rewiring reduces the diameter of the network. 	There are about $\gamma \frac{nd}2$ ``non-lattice'' edges. Varying $\gamma$ allows interpolation between the regular ring lattice ($\gamma=0$) and a network close to the Erd\"os-Renyi graph $G(n,p)$, e.g. \cite[Chapter 4]{vdH:17}, with the edge inclusion probability $p=d/(n-1)$ ($\gamma=1$). However, note that in contrast to an $G(n,p)$ in the Watts-Strogatz graph every node has at least $d/2$ neighbors. Situations with a high mean degree $d$ allow to model perturbations of the complete graph.
\end{enumerate}

With the above definition (1) becomes a special case of (2), where the number of neighbors is $N-1$. Furthermore, (2) can be considered as a special case of (3) if $\gamma=0$. 

After the construction of the underlying graph, we (re)-assign the node ids randomly.

\subsection{Faulty and malicious nodes}\label{sec:faultyMaliciousNodes}
 
\subsubsection{Positions of the faulty and malicious nodes}
We assume that there is a proportion of $q$ faulty or malicious nodes, and we call the remainder of the nodes normal or honest, respectively. There are two natural possibilities to choose faulty nodes. In the first, we choose $\lceil qn\rceil$ faulty or malicious nodes at random. This is done by performing sampling without replacement. In the second, each node becomes faulty with probability $q$ independent of all the other nodes. While the second way seems to be more natural in the modeling of faulty nodes, we choose the first method since it is more appropriate for modeling malicious nodes. In any case, the differences in the simulation outcomes are rather marginal, especially if $n$ is large.

As the assignment of the node ids in the construction of the network topology is random, we can choose the first $n_h=n-\lceil qn\rceil$ nodes as honest (normal) nodes and the remaining nodes as malicious (faulty) nodes.

We want to note that our results rely on the fact that the malicious nodes are chosen at random. An attacker that has means to interfere with the network topology might, for instance, perform eclipse attacks or use specific properties of the network topology.

\subsubsection{Initial configuration of opinions}
An opinion $s_{i}(0)\in\{0,1\}$ is assigned to all normal (honest) nodes at time $t = 0$. These opinions form the initial opinion of the nodes. 
The mean $$p_{0} = \frac1{n_h} \sum_{i=1}^{n_h} s_{i}(0)$$ is the initial proportion of $1$ opinion, where $p_{0} \in[0,1]$. 

 We initiate all nodes having an index smaller than $p_0 n_h$ with $1$, while the remainder of nodes has opinion $0$.
 
 Previous studies, e.g. \cite{GoMaMaBe:15}, used a flip-coin procedure to assign the initial opinions, that is, each node obtains its opinion by an independent Bernoulli trial in which the opinion $1$ is chosen with probability $p_{0}$ and the opinion $0$ is chosen with probability $1-p_{0}$. Due to the reassignment of node ids these two approaches essentially coincide for $n$ large. 

\subsubsection{Types of failures}\label{sec:failures}

In this paper we focus on three types of failures:

\begin{enumerate}
\item \textit{Termination failure}: 
We say the protocol suffers a termination failure if some nodes reach the maximum round $\verb?maxIt?$. 

\item \textit{Agreement failure}: 
The protocol is said to have an agreement failure if the opinion with which the normal or honest nodes terminate, is not the same for all nodes. In the majority of the cases a termination failure would also lead to an agreement failure. 

\item \textit{Integrity failure}: 
We say that the protocol suffers an integrity failure if the final opinion for all nodes is the opposite of the initial honest majority. Naturally, an agreement failure would also lead to an integrity failure. 

\end{enumerate}

The constraint on what counts as a failure may be relaxed by requiring that for at least a proportion $\epsilon$ of the honest nodes the above definition of failure occurs. However, in this paper, we take the more conservative approach and assume that $\epsilon=0$.

\subsubsection{Faulty nodes and message loss} 

Faulty nodes are assigned an opinion opposite to the initial majority of the normal nodes. We note $s^{\mathrm{n}}$ for the opinions of the normal nodes and $s^{\mathrm{f}}$ for the opinions of the faulty nodes. The dynamics are then formally defined by:
\begin{equation*}
s^{\mathrm{f}}_{i}(t) =  \left\{ \begin{array}{ll}
0, \mbox{ if } \sum_{i} s^{\mathrm{n}}_{i}(0)\geq 0.5, \\
1, \mbox{ otherwise} 
\end{array}\right. \quad \forall t\geq 1.
\end{equation*}
The update of the opinions $s^{\mathrm{n}}_{i}(t)$ of the normal nodes is done according to the protocols described in the previous section.
All faulty nodes answer queries of normal nodes about their state value but never change their own state. Since the strategy MinVS, defined in Section \hyperref[sec:maliciousNodes]{\ref{sec:maliciousNodes}}, has the same effect as the above modeling of faulty nodes, we give simulation results only for the MinVS. 

A loss of messages can severely decrease the performance and safety of the studied protocols, since a node's decision is based on less state information. We consider the model, where a query from node $i$ to node $j$ is lost with message loss probability $\mathcal{E}_{i,j}\in[0,1)$:
\begin{equation*}
s_{i,j}(t)= \left\{ \begin{array}{ll}
\star, & \mbox{ with probability } \mathcal{E}_{i,j}, \\
s_{i,j}(t) , &\mbox{ with probability } 1- \mathcal{E}_{i,j}.
\end{array}\right.
\end{equation*}
Using an indicator function we can write the number of successful queries as
\begin{equation*}
k_{i}(t)= \sum_{j\in C_{i}(t)} \mathbf{1} \{s_{i,j}(t)\in\{0,1\}\}
\end{equation*}
and the definition of $\eta$ becomes
\begin{equation*}
\eta_{i}(t+1)=\frac1{k_{i}(t)} \sum_{j\in C_{i}} s_{j}(t) \mathbf{1} \{s_{i,j}(t)\in\{0,1\}\}.
\end{equation*}

For sake of brevity and better presentation, we do not present simulation results on message loss. We content ourselves noting that performances of the FPC are in general much better under message loss than in Byzantine infrastructure. We, therefore, focus in this study on the performance of the protocol in Byzantine environment without message loss.

\subsubsection{Malicious nodes}\label{sec:maliciousNodes}
We assume that a proportion of $q\in[0,1)$ nodes are malicious and try to interfere with the protocol.
As in \cite{Po:19} we distinguish between two different kinds of adversaries: 
\begin{itemize}
\item Cautious adversary: any adversarial node must maintain the same opinion in the same round, i.e., respond the same value to all the queries it receives in that round.
\item Berserk adversary: an adversarial node may respond different opinions to different queries in the same round.
\end{itemize}

The reason for this distinction is that Berserk adversaries are much more harmful but can easily be detected if honest nodes communicate with each other. Hence, in networks with fault detection, a Berserk strategy might not be feasible for an adversary. An adversary may also choose to answer only to some nodes and to follow a semi-cautious strategy, see \cite{Po:19}. For sake of a better presentation we suppose in this paper that nodes always answers all queries.

\subsubsection*{Strategy for integrity failure.}
By monotonicity of the majority rule, the worst-case strategy for integrity failure is the initial minority vote strategy (\textbf{MinVS}): The adversary tries to turn the initial majority in transmitting the opinion of the initial minority. 
The dynamics of this strategy are exactly the same as in the setting for faulty nodes.

\subsubsection*{Cautious strategy for agreement and termination failure}
We consider the cautious strategy where the adversary transmits at time $t+1$ the opinion of the minority of the honest nodes of step $t$. We call this strategy the inverse vote strategy (\textbf{IVS}).

\subsubsection*{Berserk strategies for agreement and termination failure}

We consider a Berserk strategy, that we dub \textit{maximal variance strategy} (\textbf{MVS}). In this approach, the adversary waits until all honest nodes received opinions from all other honest nodes. The adversary then tries to subdivide the honest nodes into two equally sized groups of different opinions while trying to maximize the variance of the $\eta$-values. 

\begin{algorithm}[t]
\DontPrintSemicolon
  \KwInput{opinion}
  \KwOutput{eta}
 \For{ \upshape node in nodesHonest}{
 c[node]:= SAMPLE($\mathcal{N}\setminus$node, k, replace=FALSE)\;
 etaInter[node]:= MEAN(opinion[c[node] $\cap$ nodesHonest])\;
 kHonest[node]:= LENGTH(c[node] $\cap$ nodesHonest)\;
  completed[node] := (kHonest[node]=k)\;
  }
   \While{\upshape (LENGTH(!completed)$<$nHonest)}{
 \uIf {\upshape MEDIAN(etaInter)$>$0.5}{  
  node := ARGMIN(etaInter[!completed])\;
   opinionAdv[node]:=0
   }
   \Else {
   node := ARGMAX(etaInter[!completed])\;
   opinionAdv[node]:=1
    }
    etaInter[node]:= (etaInter[node] *  kHonest[node] \;
    \quad\quad\quad\quad\quad\quad\quad\quad+ opinionAdv[node]*(k-kHonest[node]))/k\;  
   completed[node] := TRUE\;
   }
   eta := etaInter
\caption{Adapted Berserk Strategy: the maximal variance strategy (MVS).}
\label{alg:MVS}
\end{algorithm}

In order to maximize the variance, the adversary requires knowledge over the $\eta$-values of the honest nodes at any given time. The adversary then answers queries of undecided nodes in such a way that the variance of the $\eta$'s is maximized by keeping the median of the $\eta$'s close to $0.5$, see Algorithm \hyperref[alg:MVS]{\ref{alg:MVS}} for a formal definition.

\section{Performance evaluation}\label{sec:simulation}

\subsection{Performance measures}\label{sec:performanceMeasures}

In the following, we define the metrics that we employ to analyze the performance of our protocol. The first three definitions are rates that are based on the failures defined in Section \ref{sec:failures}. 

\subsubsection{Termination rate}
We are interested in the rate at which the protocol terminates for all nodes before the maximum round $\verb?maxIt?$. For any given simulation run we say the termination is 1 if all nodes conclude before $\verb?maxIt?$, while it is 0 otherwise. The rate is determined by averaging over the termination values for several simulation runs. 

\subsubsection{Agreement rate}
We study the performance of the consensus protocols in terms of their agreement rate (also known as convergence rate, see Section \hyperref[sec:binaryconsensus]{\ref{sec:binaryconsensus}}) that is the fraction of initial configurations (or simulations) resulting in successful agreement. If agreement is achieved we assign the value 1, or 0 otherwise. Similarly to the termination rate the agreement rate is obtained by averaging over several simulation runs. 

\subsubsection{Integrity Rate}
We say integrity is achieved, if the opinion of the initial honest majority is preserved after the protocol concludes, i.e. the protocol finishes with 1 if $p_0\geq0.5$ or 0 if $p_0<0.5$. In the first case, we also speak of $1$-integrity and in the second case we speak of $0$-integrity.

If integrity is achieved we assign the value 1, or 0 otherwise. The integrity rate is obtained by averaging over several simulation runs.

\subsubsection{Maximal time to termination $\overline{T}_{max}$} Average time for the protocol to terminate across the entire network, i.e.\ the time until the last node terminates the protocol. 

\subsubsection{Mean time to termination $\overline{T}_{mean}$} Average time for the protocol to terminate considering all nodes. Since each node queries $k$ nodes each round, the average total number of messages sent, i.e.\ the message complexity, is given by 
\begin{equation*}
  C=  \overline{T}_{mean} k n
\end{equation*}

\subsection{Simulation methodology and parameters}

In this section, we analyze the FPC protocol based on the network topologies and malicious actors defined in the previous sections. For most of our analysis, we assume the default parameter set in Table \hyperref[tab:defaultparams]{\ref{tab:defaultparams}} and any deviations from these values are mentioned explicitly.
\\

\begin{minipage}{\textwidth}
\center
\captionof{table}{Default simulation parameters}
\begin{tabular}{llc}
\hline
Parameter & & Value\\ \hline\hline
$n$ & Number of nodes & 1000\\ \hline
$\tau$ & Initial threshold & 2/3 \\\hline
$k$& Quorum size & 21 \\\hline
$\beta$ & Lower random threshold bound& 0.3\\\hline
$\el$  & Final consecutive round & 10 \\ \hline
$\verb?maxIt?$ & Max termination round & 100 \\ \hline
$q$ & Proportion of adversarial nodes & 0.1 \\\hline
\hline
\newline
\end{tabular}
\label{tab:defaultparams}
\end{minipage}\hfill

Figures shown in this section are obtained from data for which each data point is the average over a sample of at least 10,000 simulation runs. The high sample size per data point ensures that the standard error of the mean value is less than 1\%.

\subsection{FPC with a partial network view}\label{sec:sim:network-partitioning}

We study the FPC-protocol on non-complete graphs and in a setting without an adversary. We utilize the Watts-Strogatz graph and the associated parameters $\gamma$ and $\delta$ that have been described in Section \hyperref[sec:networktopology]{\ref{sec:networktopology}} to analyze the effects of different network topologies on the performance of FPC. 

\begin{figure}[h]
\begin{center}
\begin{minipage}{0.45\textwidth}
   \hspace{-1cm}
    \includegraphics[width=1.3\textwidth,trim={0 0 0 1cm},clip]{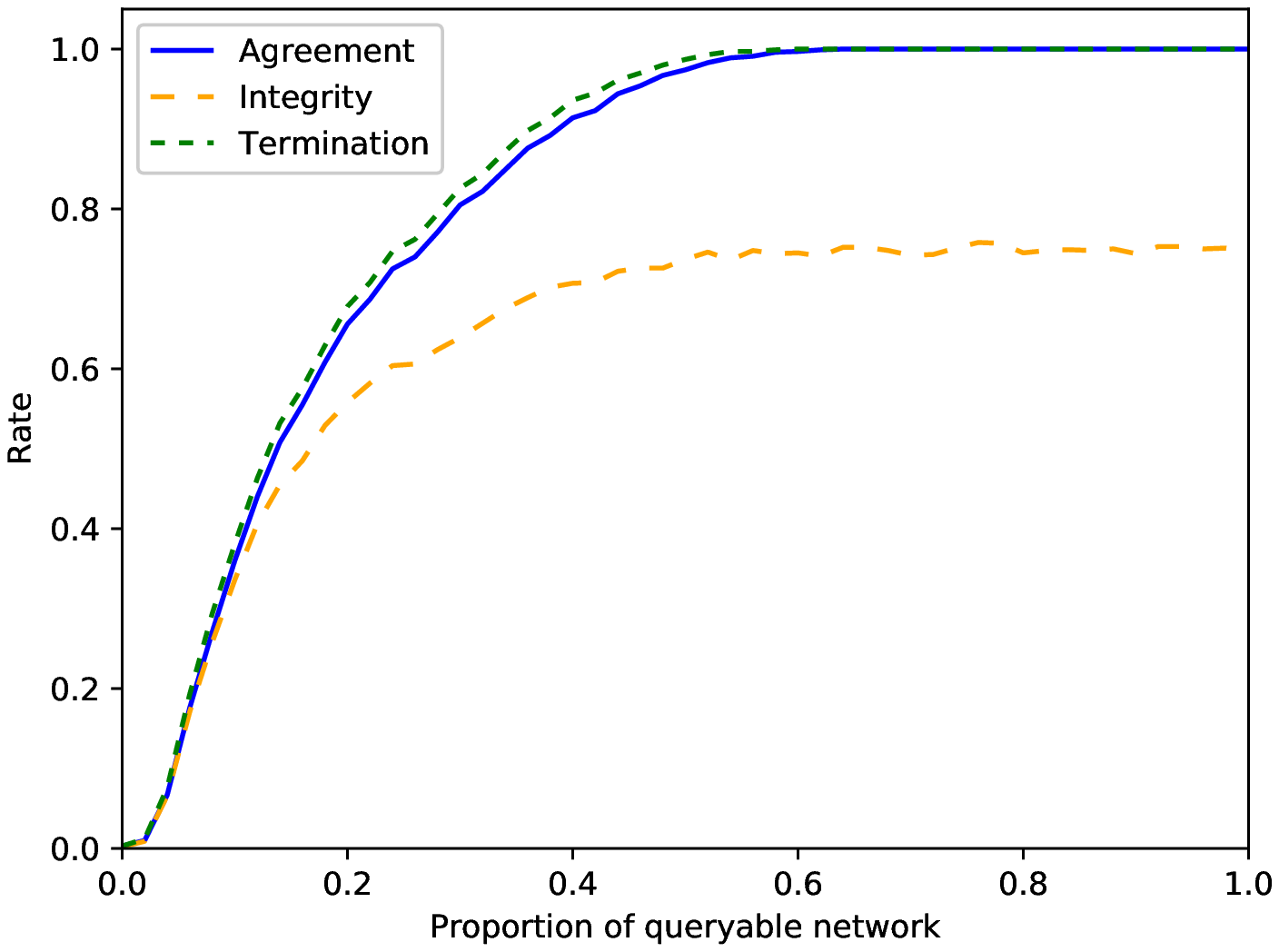}
    \vspace{-0.8cm}
  \captionof{figure}{\small Termination, agreement and integrity rate as a function of the proportion $\delta$ of the network that is queryable by a node, for a ring lattice graph.}
  \label{fig:eta-TAI}
\end{minipage}\hfill
\begin{minipage}{.45\textwidth}
   \hspace{-1.2cm}
    \includegraphics[width=1.3\textwidth,trim={0 0 0 1cm},clip]{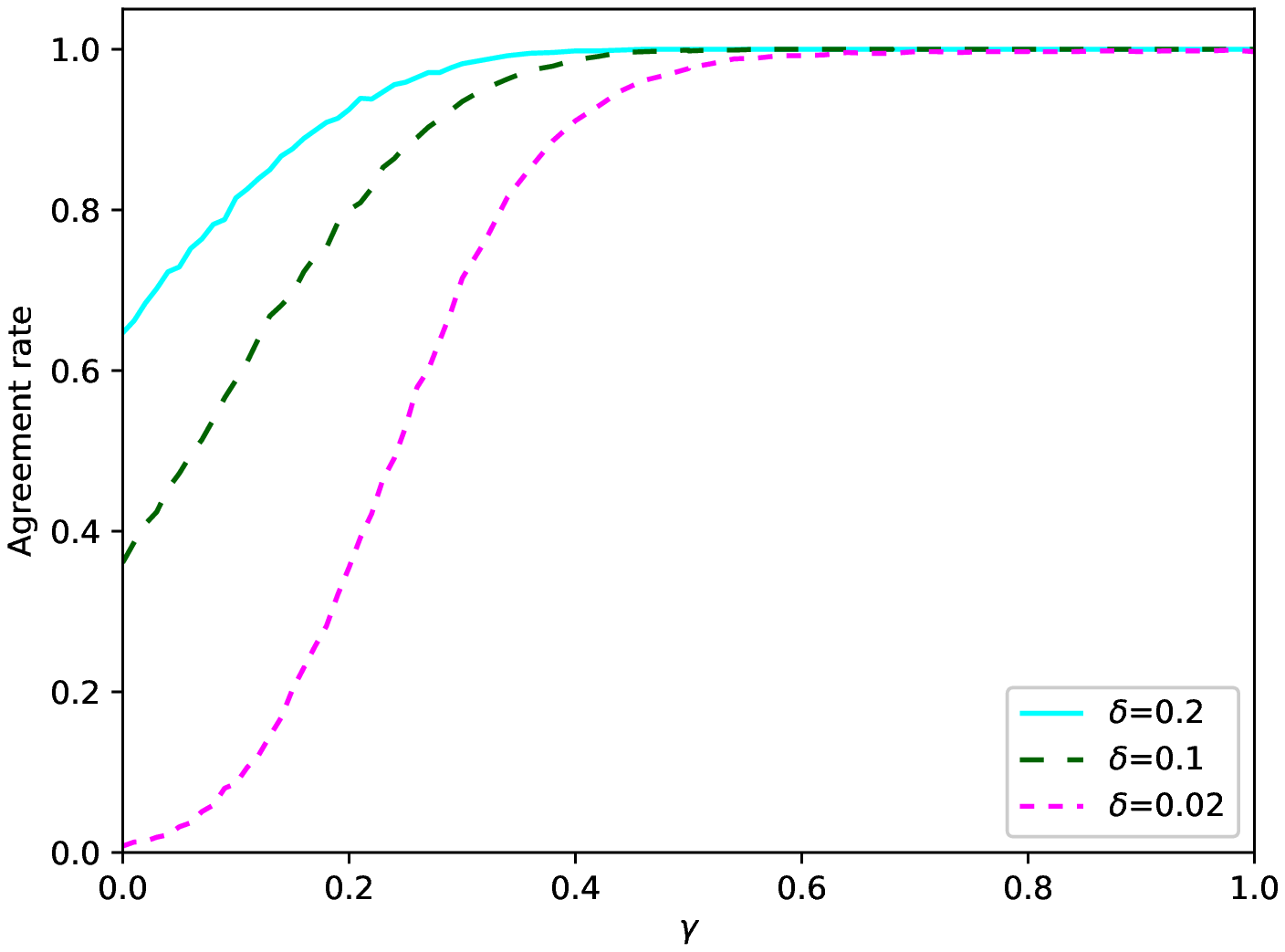}
    \vspace{-0.8cm}
    \captionof{figure}{\small Agreement rate as a function of the rewiring probability $\gamma$ for several values of $\delta$.\newline}
    \label{fig:gamma-A}
\end{minipage}\hfill
\end{center}
\end{figure}

In Fig. \hyperref[fig:eta-TAI]{\ref{fig:eta-TAI}} we show the performance of the protocol on a ring lattice, which is equivalent to a Watts-Strogatz graph with rewiring probability $\gamma=0$. Furthermore, we set the proportion of initial 1-opinions to $p_0=\tau$. This corresponds to the case where we expect the highest integrity failure. It can be seen that if the ring lattice nodes do not have a sufficient amount of neighbors, termination and hence agreement cannot be achieved. This is caused by the random distribution of initial opinions which results in locally differing 0- or 1-majorities. 

The performance at values of $\delta<0.5$ can be improved by rewiring some of the connections of the nodes, according to the method described in Section \hyperref[sec:networktopology]{\ref{sec:networktopology}}. Fig. \hyperref[fig:gamma-A]{\ref{fig:gamma-A}} shows the variation of the agreement rate with the rewiring probability $\gamma$. We see that even if only a small proportion of the network can be queried, the protocol can come to agreement in an honest setting if a sufficient amount of nodes are rewired. For example, if the queryable proportion of the network is $\delta=0.1$, it is sufficient if $\approx30\%$ of the nodes get rewired. 

For sake of simplicity and to enable comparison with theoretical results we assume a complete graph for the network in the following sections. However, as we have shown in this section the partial network view has noticeable effects on the termination and agreement rate. We will, therefore, further investigate on the effects of a partial network view in Section \hyperref[sec:sim:TermAgree]{\ref{sec:sim:TermAgree}}. 

\subsection{Integrity failure} \label{sec:sim:Int}

An adversary may attempt to change the initial majority-opinion in the network. If successful we say that the protocol suffers an integrity failure. We distinguish two types of integrity failures. Firstly the adversary may attempt to change the opinion for which initially the majority opinion is 0, and we denote a successful inversion of the opinion as a 0-integrity failure. Secondly, if initially, the majority of nodes  would vote 1 an integrity failure would be called 1-integrity failure. 

The most effective strategy for the adversary to achieve an integrity failure is to continuously vote the opposite of the initial majority. We will therefore, focus on this strategy in this Section. 

\subsubsection{\textbf{The 0-integrity failure}}\label{sec:sim:0Int}

In order to assess the capability of the protocol to support an asymmetric behavior as described in the beginning of  Section \ref{sec:protocols}, we investigate the likelihood of a 0-integrity failure with the strategy MinVS defined in Section \hyperref[sec:maliciousNodes]{\ref{sec:maliciousNodes}}. 

As discussed in Section \hyperref[sec:protocols]{\ref{sec:protocols}} the asymmetry is enabled through the introduction of the initial threshold $\tau$. Since this threshold is introduced to support the asymmetric importance of the two integrity failures, we investigate the protocol in a worst-case scenario, where 51\% of the honest nodes have opinion 0, i.e. $p_0=0.49$.

Fig. \hyperref[fig:a-I0-varyq]{\ref{fig:a-I0-varyq}} shows the 0-integrity rate with the initial threshold $\tau$ for several values of proportions of nodes controlled by the adversary $q$. The fact that the integrity rate resembles a staircase function can be explained by the fact that $\eta$'s can only take a finite number of values. As expected the integrity rate increases with $\tau$. However, for the default parameter set (see Table \hyperref[tab:defaultparams]{\ref{tab:defaultparams}}) the integrity rate is limited below 1 for $q=0.2$. On the other hand, by comparing Fig. \hyperref[fig:a-I0-varyq]{\ref{fig:a-I0-varyq}} and Fig. \hyperref[fig:a-I1-varyq]{\ref{fig:a-I1-varyq}} it can be seen that for $q=0.1$ a range for the initial threshold exists for which both 0-Integrity and 1-integrity is fully achieved.

\begin{center}
\begin{minipage}{0.45\textwidth}
  \hspace{-1cm}
  \includegraphics[width=1.3\textwidth,trim={0 0 0 1cm},clip]{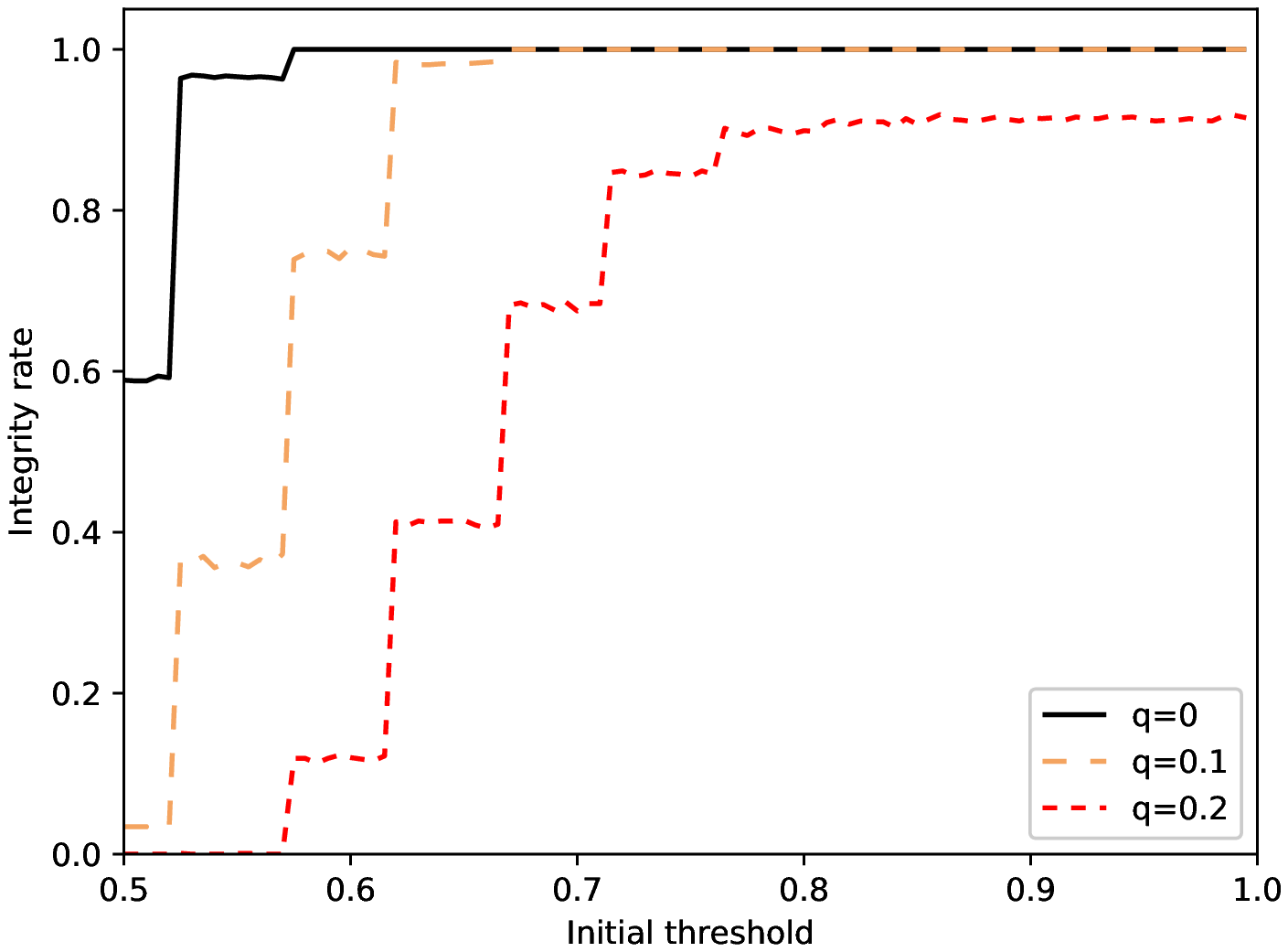}
  \vspace{-0.8cm}
  \captionof{figure}{\small 0-Integrity rate as a function of the initial threshold $\tau$ with $p_0=0.49$.  }
  \label{fig:a-I0-varyq}
\end{minipage}\hfill
\begin{minipage}{.45\textwidth}
  \hspace{-1.2cm}
  \includegraphics[width=1.3\textwidth,trim={0 0 0 1cm},clip]{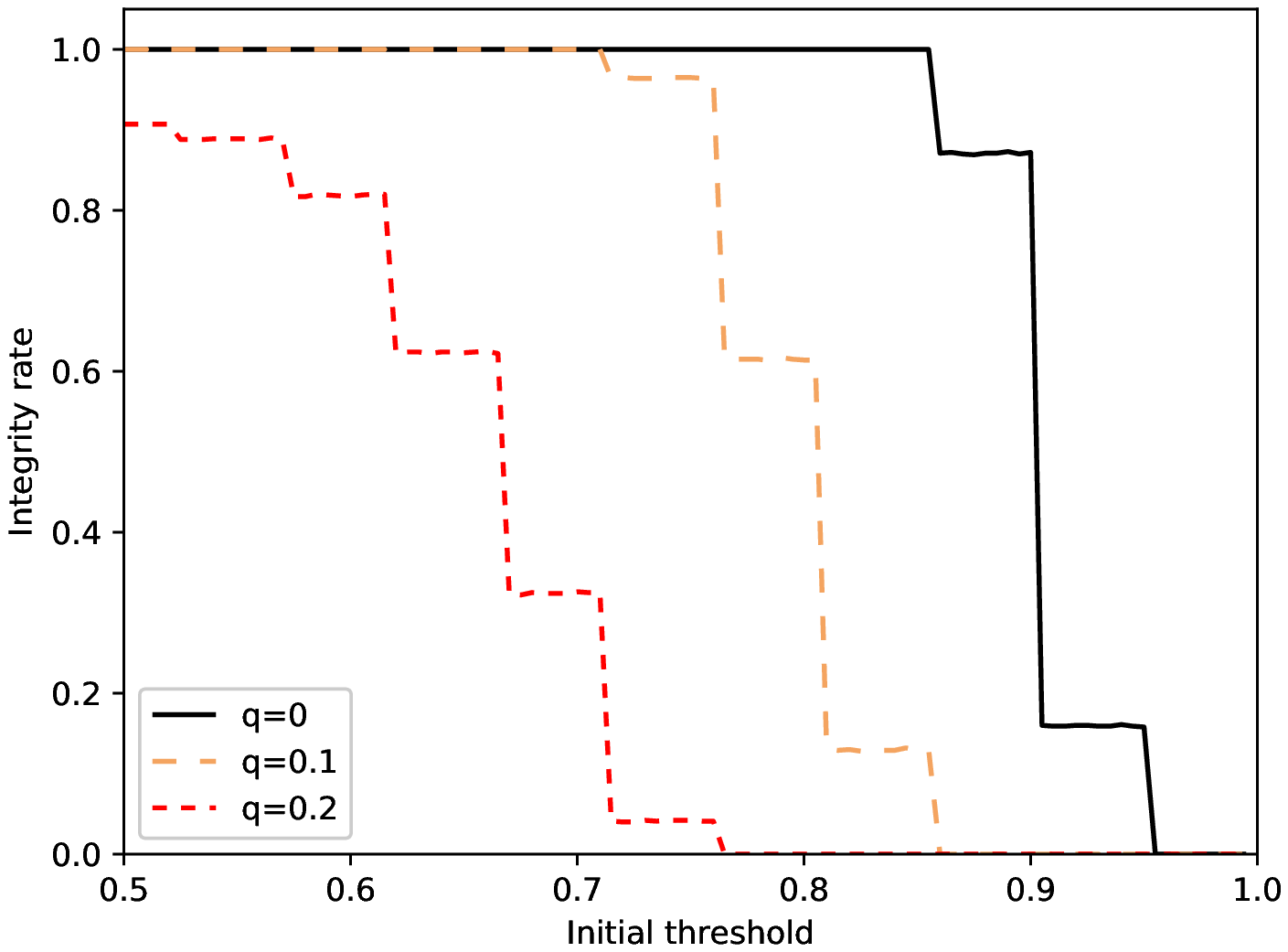}
  \vspace{-0.8cm}
  \captionof{figure}{\small 1-Integrity rate as a function of the initial threshold $\tau$ with $p_0=0.9$.}
  \label{fig:a-I1-varyq}
\end{minipage}\hfill
\end{center}

\subsubsection{\textbf{The 1-integrity failure}}\label{sec:sim:1Int}

In order to protect against a 0-integrity failure the threshold has to be above $0.5$, i.e. $\tau>0.5$. Albeit a high value for $\tau$ is a good protection mechanism against a 0-integrity failure it can increase the possibility for a 1-integrity failure. The initial threshold should be, therefore, tuned carefully. The integrity rate can be particularly important if the initial 1-opinion is a supermajority, i.e. it is close to 1. Hence in this section we set $p_0=0.9$. As in the previous section we employ the MinVS. 

Fig.\ \hyperref[fig:a-I1-varyq]{\ref{fig:a-I1-varyq}} shows the integrity rate as a function of the initial threshold $\tau$. The termination and agreement rate is not presented because it is either equal or close to one. We see that if $\tau$ is selected too high, 1-integrity is not guaranteed. 

\subsubsection*{Minimum required consecutive rounds} 
\begin{figure}[h]
\begin{center}
\begin{minipage}{.45\textwidth}
   \hspace{-1cm}
    \includegraphics[width=1.3\textwidth,trim={0 0 0 1cm},clip]{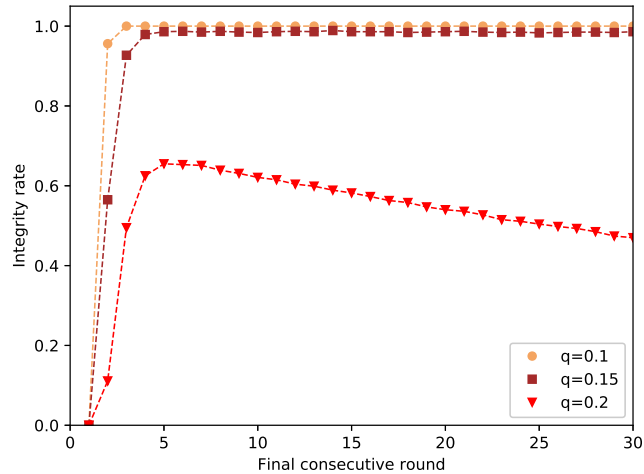}
    \vspace{-0.8cm}
   \captionof{figure}{\small 1-Integrity rate as a function of the final consecutive round $\el$}
   \label{fig:l-I}
\end{minipage}\hfill
\end{center}
\end{figure}

In order to achieve a low message complexity, the final consecutive round $\el$ should be chosen low for the protocol to terminate quickly. Indeed as can be seen from Fig. \hyperref[fig:l-I]{\ref{fig:l-I}} the protocol performs well even for a small $\el$ if $q=0.1$. However if the adversary controls a larger proportion of the network the capacity of the protocol to perform well at lower values decreases, as can be seen for $\el<6$. Furthermore, if the adversary controls more than a certain amount of nodes a higher $\el$ does not always increase the integrity rate, as can be seen for $q=0.2$. 

\subsubsection*{Quorum size} 
\begin{figure}{h}
\begin{center}
\begin{minipage}{0.45\textwidth}
   \hspace{-1cm}
   \includegraphics[width=1.3\textwidth,trim={0 0 0 1cm},clip]{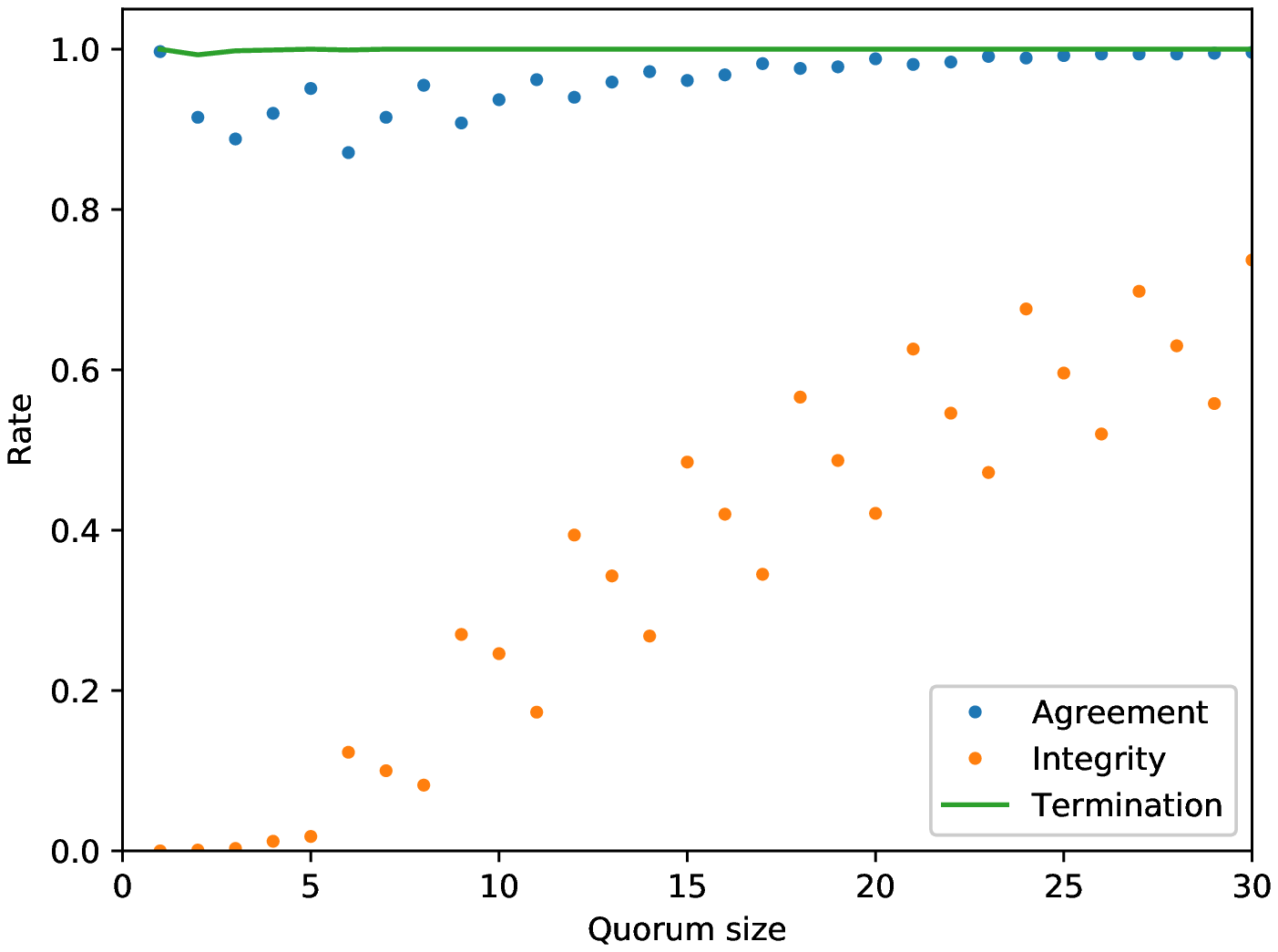}
    \vspace{-0.8cm}
    \captionof{figure}{\small Termination and integrity rate as a function of the variation of the number of queried nodes $k$. }
    \label{fig:k-TAI-1}
\end{minipage}\hfill
\begin{minipage}{.45\textwidth}
   \hspace{-1.2cm}
  \includegraphics[width=1.3\textwidth,trim={0 0 0 1cm},clip]{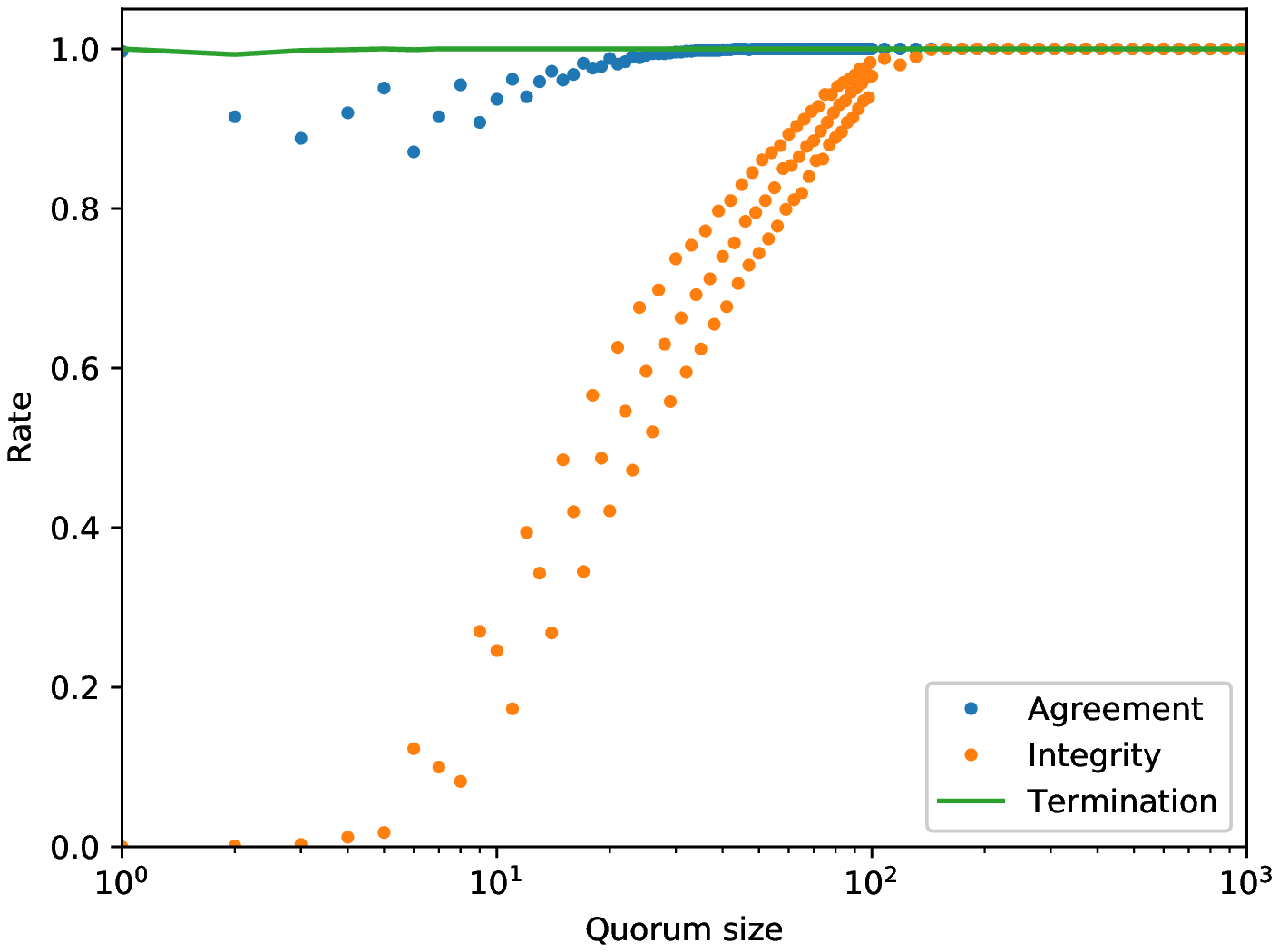}
  \vspace{-0.8cm}
    \captionof{figure}{\small Termination and integrity rate as a function of the variation of the number of queried nodes $k$.}
   \label{fig:k-TAI-2}
\end{minipage}\hfill
\end{center}
\end{figure}

The protocol can be improved by increasing the number of queried nodes $k$, as can be seen from Figs. \hyperref[fig:k-TAI-1]{\ref{fig:k-TAI-1}}-\hyperref[fig:k-TAI-2]{\ref{fig:k-TAI-2}}. However, this comes at the expense of an increased message complexity. Similarly to the analysis with varied $\tau$, see Fig.\ \hyperref[fig:a-I1-varyq]{\ref{fig:a-I1-varyq}}, we can observe a discrete dependency of the integrity and agreement rate on $k$. Again this is caused by the discreteness of the possible $\eta$-values and their relation to the initial threshold $\tau$.

\subsubsection*{Scalability}
\begin{figure}[h]
\begin{center}
\begin{minipage}{0.45\textwidth}
    \hspace{-1cm}
    \includegraphics[width=1.3\textwidth,trim={0 0 0 1cm},clip]{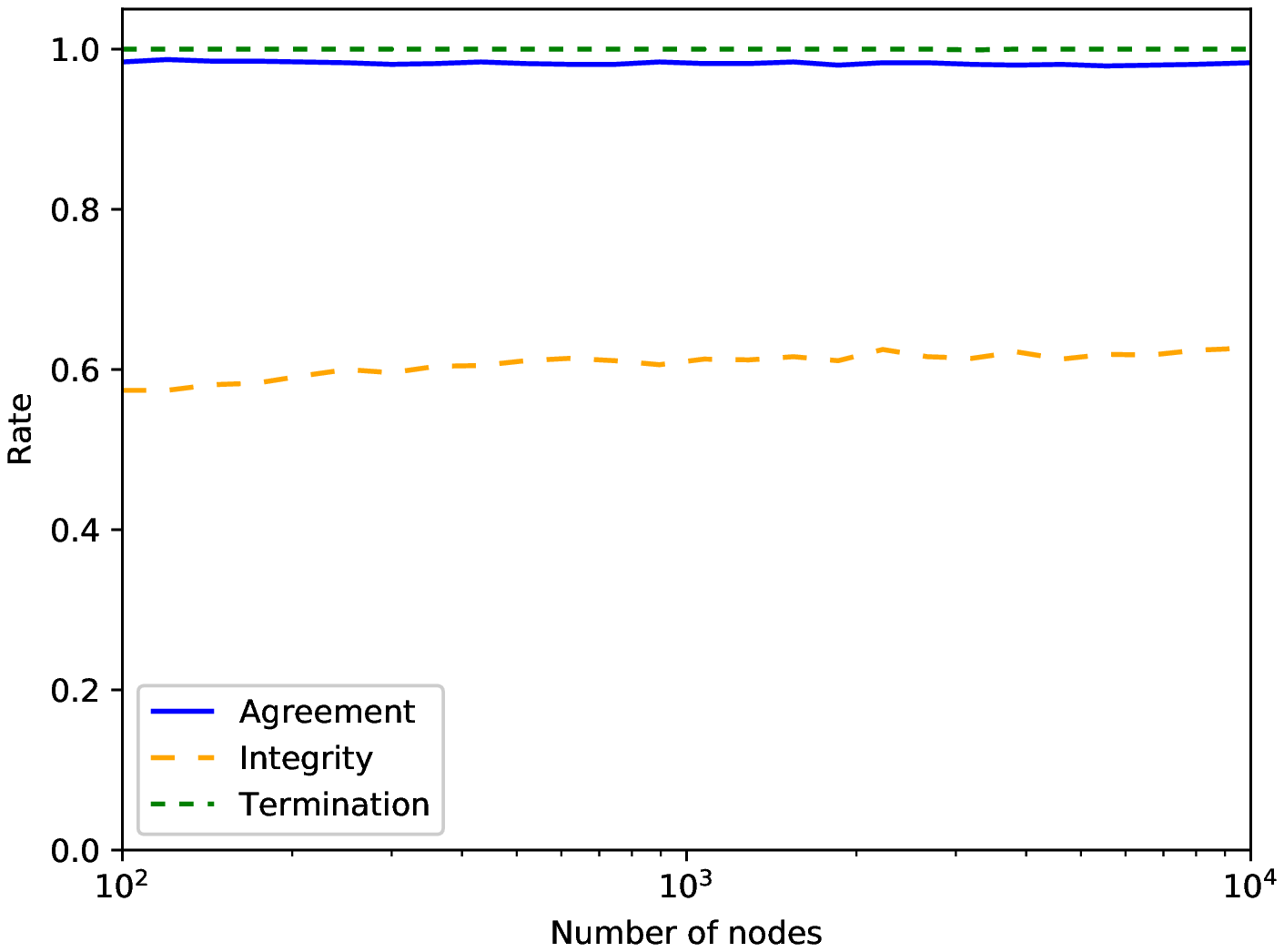}
    \vspace{-0.8cm}
    \captionof{figure}{\small Termination-, Agreement- and Integrity-Rate as a function of the number of nodes $N$ for $q=0.2$. }
    \label{fig:N-TAI}
\end{minipage}\hfill
\begin{minipage}{.45\textwidth}
  \hspace{-1.2cm}
  \includegraphics[width=1.3\textwidth,trim={0 0 0 1cm},clip]{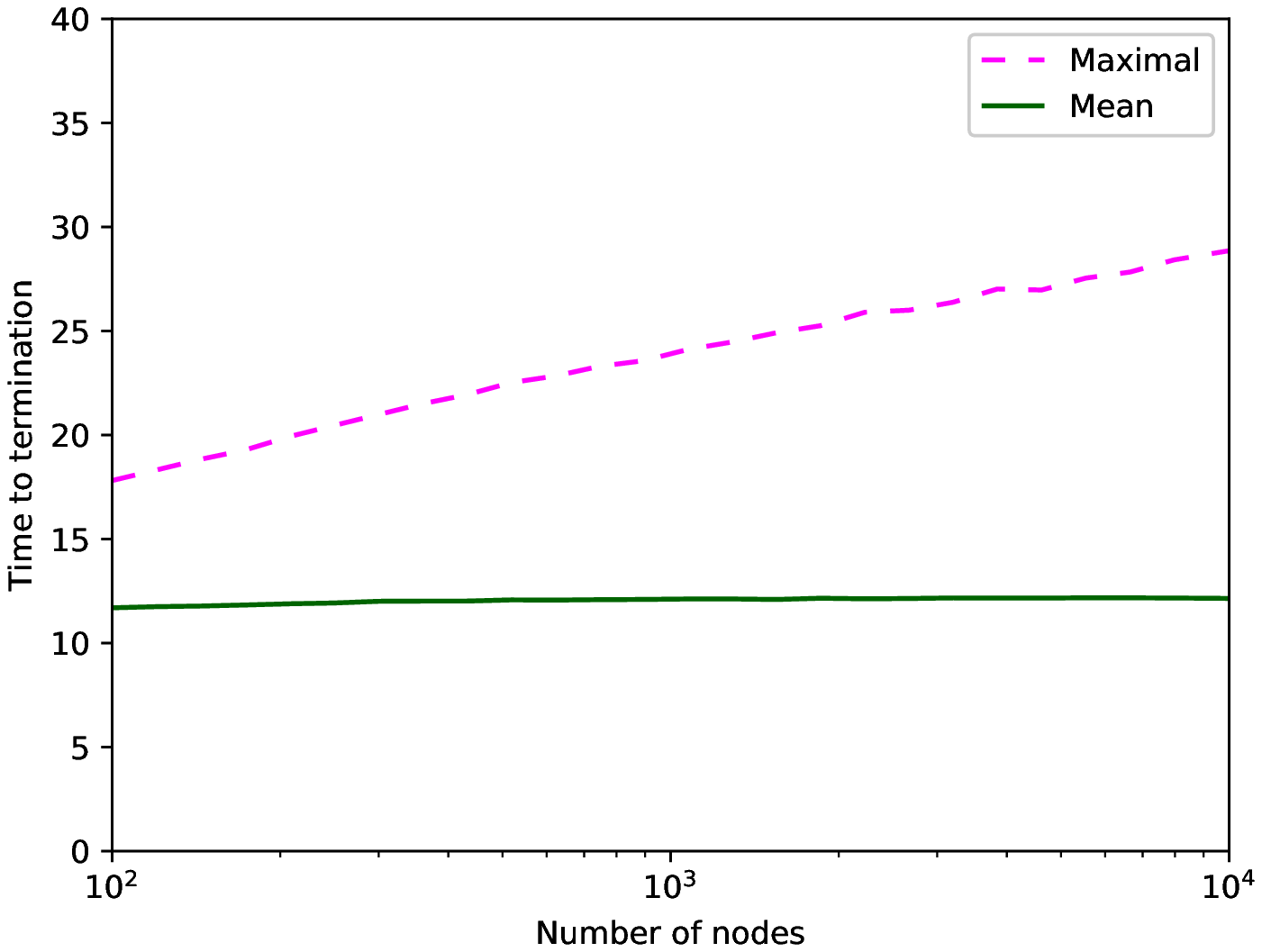}
  \vspace{-0.8cm}
  \captionof{figure}{\small Time to termination as a function of the number of nodes $N$ for $q=0.2$.}
  \label{fig:N-rounds}
\end{minipage}\hfill
\end{center}
\end{figure}

The performance of the protocol depends little on the overall size of the network for a given parameter setting. In order to display this, we keep the parameter setting to $k=21$, which leads to an integrity rate below 1. However, as indicated above this can be resolved by increasing $k$. As can be seen from Fig.\ \hyperref[fig:N-TAI]{\ref{fig:N-TAI}} the protocol achieves a similar performance as the number of nodes increases. Fig.\ \hyperref[fig:N-TAI]{\ref{fig:N-rounds}} shows that although the maximal time to termination $\overline{T}_{max}$ increases, the meantime to termination $\overline{T}_{mean}$ remains almost constant and hence the total message complexity is essentially linear in $n$, i.e. increases almost $O(n)$.

\subsubsection*{Initial mean opinion}
Generally, the protocol performs best if all of the honest nodes share the same opinion. Fig.\ \hyperref[fig:vary-qp0]{\ref{fig:vary-qp0}} shows a ``heat map'' of the integrity rate with the mean honest initial opinion $p_0$ and the proportion of adversary nodes $q$. It can be seen, that if more than 90\% of the honest nodes share the same opinion, the protocol can still withstand a 15\% attack.

\subsubsection*{Randomization of the threshold}
\begin{figure}[t]
\begin{minipage}{0.45\textwidth}
  \hspace{-1cm}
  \includegraphics[width=1.3\textwidth,trim={0 0 0 0.5cm},clip]{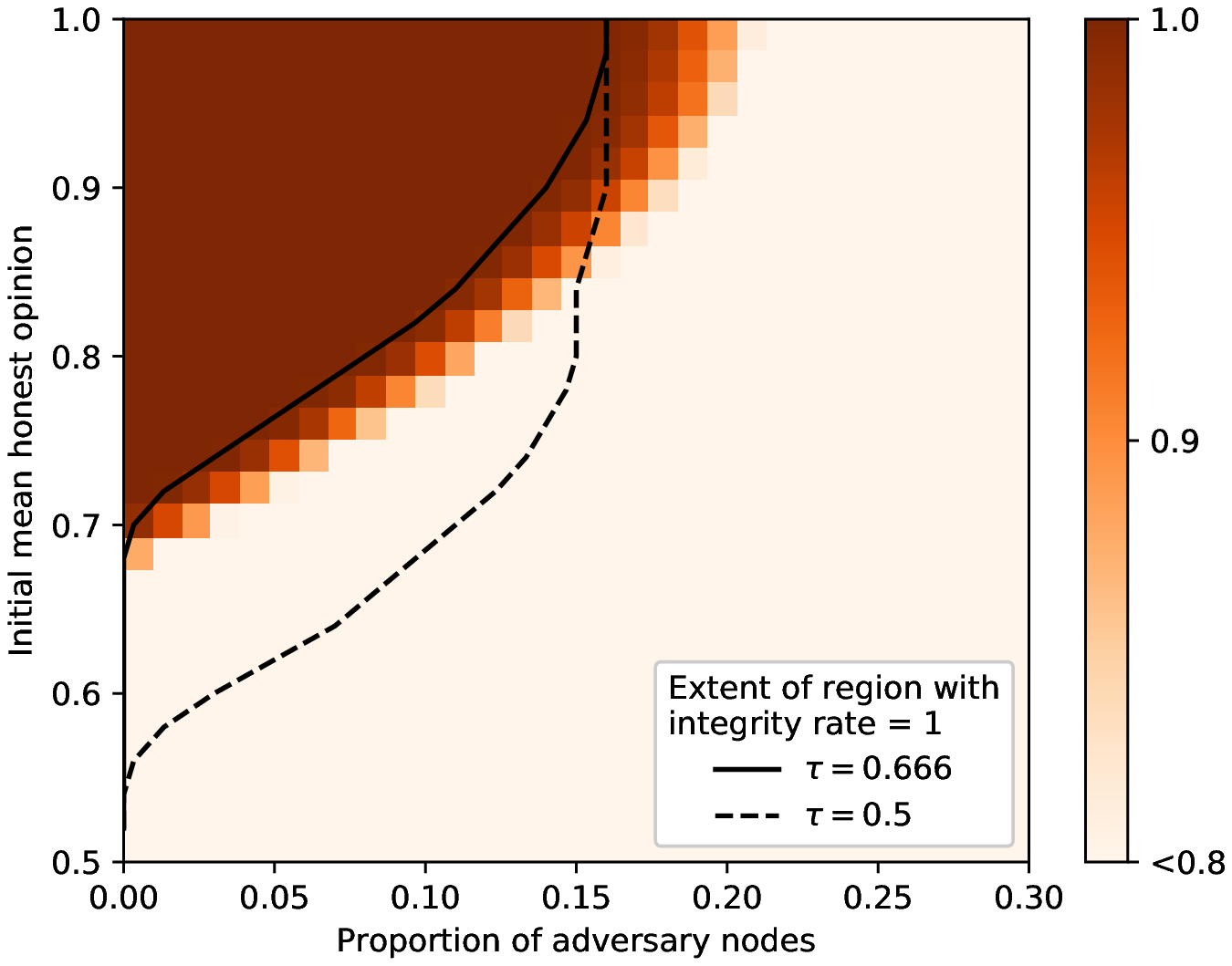}
  \vspace{-0.8cm}
  \captionof{figure}{\small Integrity rate as a function of $q$ and $p_0$.}
  \label{fig:vary-qp0}
\end{minipage}\hfill
\begin{minipage}{.45\textwidth}
  \hspace{-1.2cm}
  \includegraphics[width=1.3\textwidth,trim={0 0 0 0.5cm},clip]{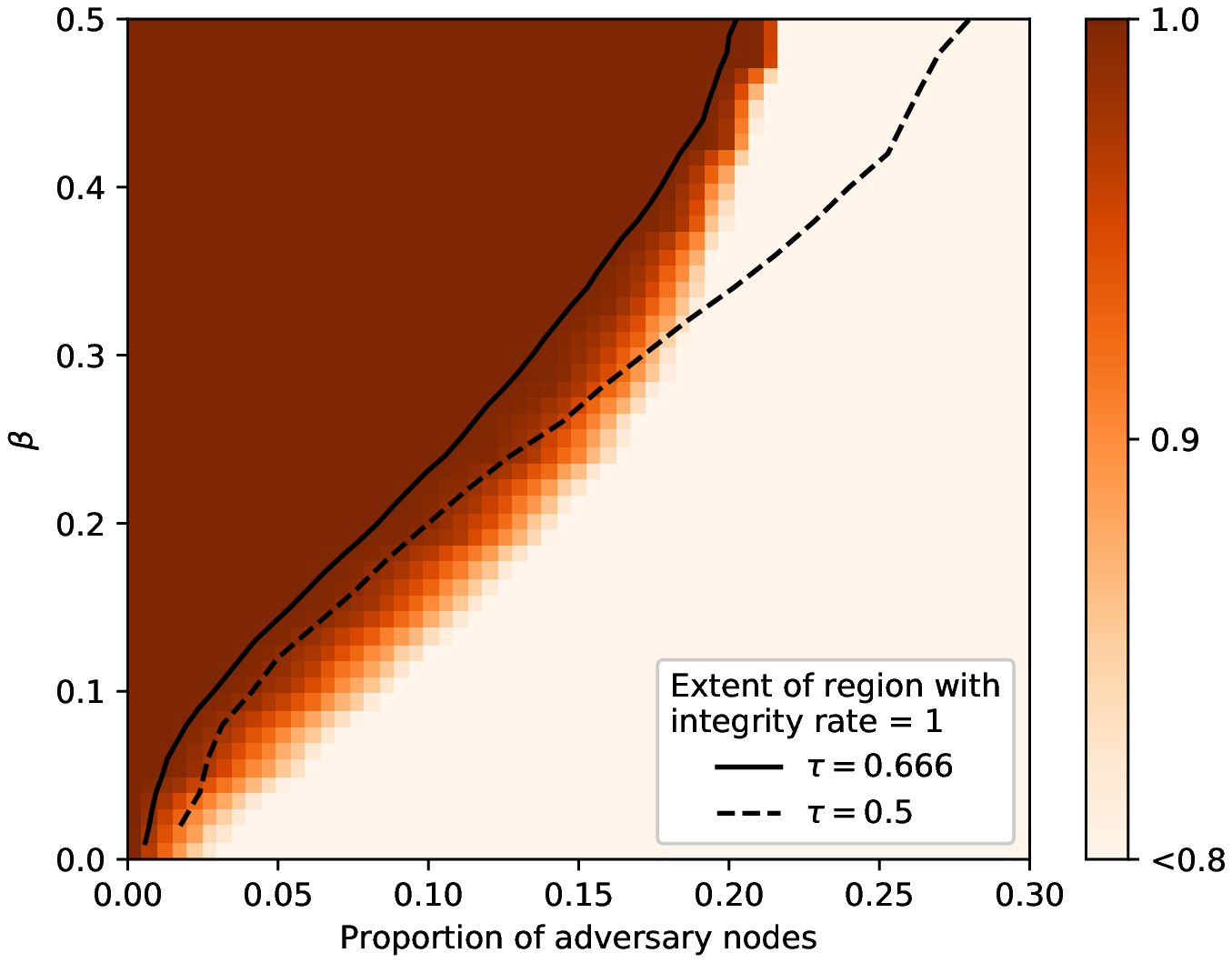}
  \vspace{-0.8cm}
  \captionof{figure}{\small Integrity rate as a function of $q$ and $\beta$. }
  \label{fig:vary-qbeta}
\end{minipage}\hfill
\end{figure}

Fig.\ \hyperref[fig:vary-qbeta]{\ref{fig:vary-qbeta}} shows that the randomness of the thresholds has some impact on the performance of the protocol in terms of integrity rate. However as long as the randomness is not above a certain threshold, i.e. $\beta\geq0.3$, the protocol can withstand an adversary proportion of about 15\%.

We also analyze the scenario where the 0-integrity failure described in Section \hyperref[sec:sim:0Int]{\ref{sec:sim:0Int}} is of less concern. This, for example, may be the case if we are certain that the initial opinion is only in one of the two scenarios: the mean initial opinion is either a super-minority or a super-majority, i.e. $p_0\ll0.5$ or $p_0\gg0.5$, respectively. Under these circumstances it would not be necessary to provide an initial threshold $\tau>0.5$. Figs. \hyperref[fig:vary-qp0]{\ref{fig:vary-qp0}}-\hyperref[fig:vary-qbeta]{\ref{fig:vary-qbeta}} show the extend of the region for which the integrity rate is one, if $\tau=0.666$ and $\tau=0.5$. From Fig.\ \hyperref[fig:vary-qp0]{\ref{fig:vary-qp0}} it can be seen that this region is increased, in particular for lower $p_0$. Similarly, Fig.\ \hyperref[fig:vary-qbeta]{\ref{fig:vary-qbeta}} shows that a integrity rate of 1 is achieved for  values of $\beta$ close to 0.5.

\subsection{Termination and agreement failure}\label{sec:sim:TermAgree}
In this section, we investigate the worst-case scenario for termination and agreement failure. An initial average opinion $p_0$ close to the initial threshold $\tau$ enables a potential attacker the most to achieve an agreement or termination failure. For instance, the protocol of a simple majority rule, see Section \hyperref[sec:rmc]{\ref{sec:rmc}}, may be prevented from terminating already with a small percentage of malicious nodes. For this reason, we set $p_0=\tau$   in this section.

\subsubsection*{Intermediate opinion distribution} 
\begin{figure}[h]
\hspace{1.2cm}
\begin{minipage}{.85\textwidth}
  \includegraphics[width=1.\textwidth,trim={0 0 0 0.cm},clip]{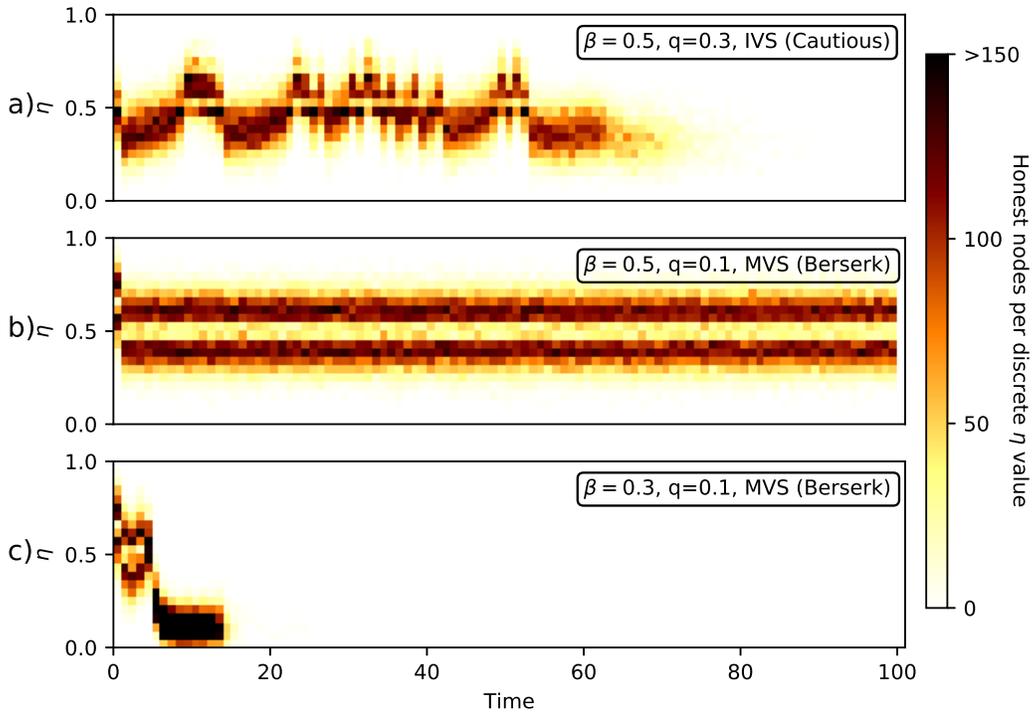}
  \vspace{-0.8cm}
  \captionof{figure}{\small Evolution of the number of undecided nodes for a given $\eta$-value.}
  \label{fig:eta-heatmap2}
\end{minipage}\hfill
\end{figure}

Fig.\ \hyperref[fig:eta-heatmap2]{\ref{fig:eta-heatmap2}} shows the development of the discrete $\eta$'s distribution for the following three scenarios for a single simulation run: 
\begin{itemize}
   \item[a)] Firstly we employ the cautious strategy IVS, see Section \hyperref[sec:maliciousNodes]{\ref{sec:maliciousNodes}}, against the protocol without randomness, i.e. $\beta=0.5$ and $q=0.3$. It can be seen that the strategy is able to keep the $\eta$ distribution close to $0.5$ for several rounds.  Due to the randomness introduced by the quorum selection the adversary ultimately fails to prevent the protocol from terminating. It should be noted that the shown simulation results for this case represent a worst-case scenario, and that for most simulation runs the final termination round remained below $t=30$. 
    
    \item[b)] In the second figure we apply the Berserk strategy MVS, see Section \hyperref[sec:maliciousNodes]{\ref{sec:maliciousNodes}} with no random threshold and the much lower adversary proportion $q=0.1$. We see that the strategy succeeds in splitting the nodes into two groups and that the protocol can be prevented from terminating. 
    
    \item[c)] In the third scenario randomness is added; i.e.\ $\beta=0.3$. 
   Due to the random threshold within a short time (less than round 5) a supermajority of the nodes' opinions is formed, i.e. the opinions converge to one value, in this case 0. After the convergence, the adversary still manages to affect the $\eta's$ of some of the nodes, as can be noticed through the formation of the small group in proximity of 0. However, since $\beta$ is large enough, the strategy fails to flip a large enough amount of node's opinion. Ultimately at about round $\el=10$ the nodes start to finalize and the protocol terminates shortly after.
\end{itemize}

\subsubsection*{Comparison of adversarial strategies} 
Fig.\ \hyperref[fig:q-TA]{\ref{fig:q-TA}} shows the termination and agreement rate with the proportion of adversarial nodes for several adversarial strategies. As can be seen with the introduction of the random thresholds the protocol is resilient against the investigated attacks in terms of termination and agreement failure for a large proportion of adversarial nodes. It can also be seen that the Berserk strategy MVS is significantly more severe than IVS.

We assess the resilience against a Berserk adversary given each node only has a partial view of the network, by employing a Watts-Strogatz graph, see Sections \hyperref[sec:networktopology]{\ref{sec:networktopology}} and \hyperref[sec:sim:network-partitioning]{\ref{sec:sim:network-partitioning}}. Fig.\ \hyperref[fig:delta-ATI]{\ref{fig:delta-ATI}} shows the termination and agreement rate with $\delta$. It can be seen, that individual nodes do not require a complete network view in order to come to agreement. Analogous to the analysis for Fig.\ \hyperref[fig:gamma-A]{\ref{fig:gamma-A}}, if the network view is partially randomized, i.e. $\gamma>0$, a smaller network view may be sufficient. Again MVS appears more severe than IVS. Therefore, for the remainder of this section, we will focus on the more troublesome MVS.

\begin{center}
\begin{minipage}{0.45\textwidth}
   \hspace{-1cm}
   \includegraphics[width=1.3\textwidth,trim={0 0 0 0.5cm},clip]{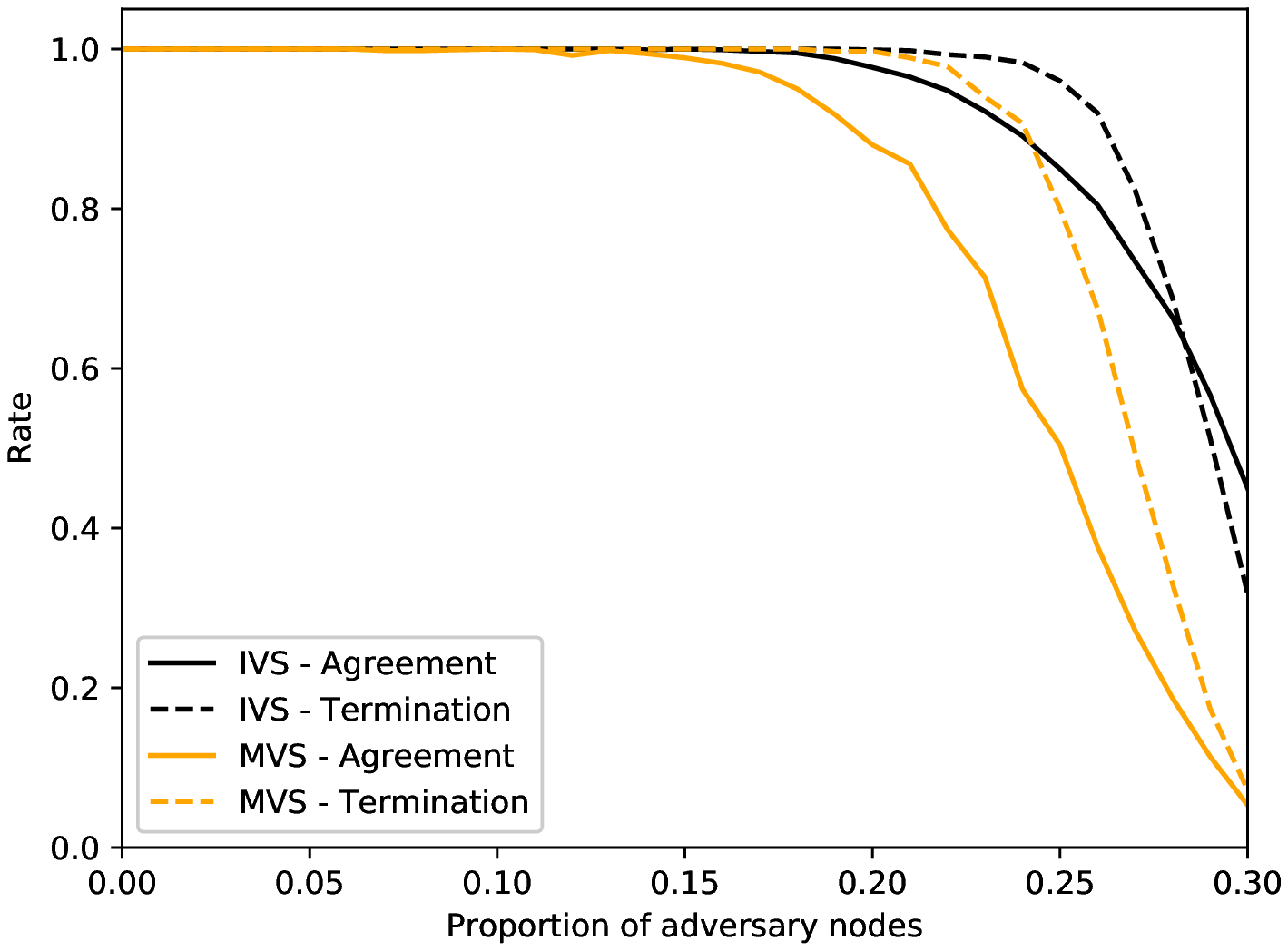}
   \vspace{-0.8cm}
  \captionof{figure}{\small Termination and Agreement rate as a function of $q$.}
  \label{fig:q-TA}
\end{minipage}\hfill
\begin{minipage}{.45\textwidth}
   \hspace{-1.2cm}
  \includegraphics[width=1.3\textwidth,trim={0 0 0 0.5cm},clip]{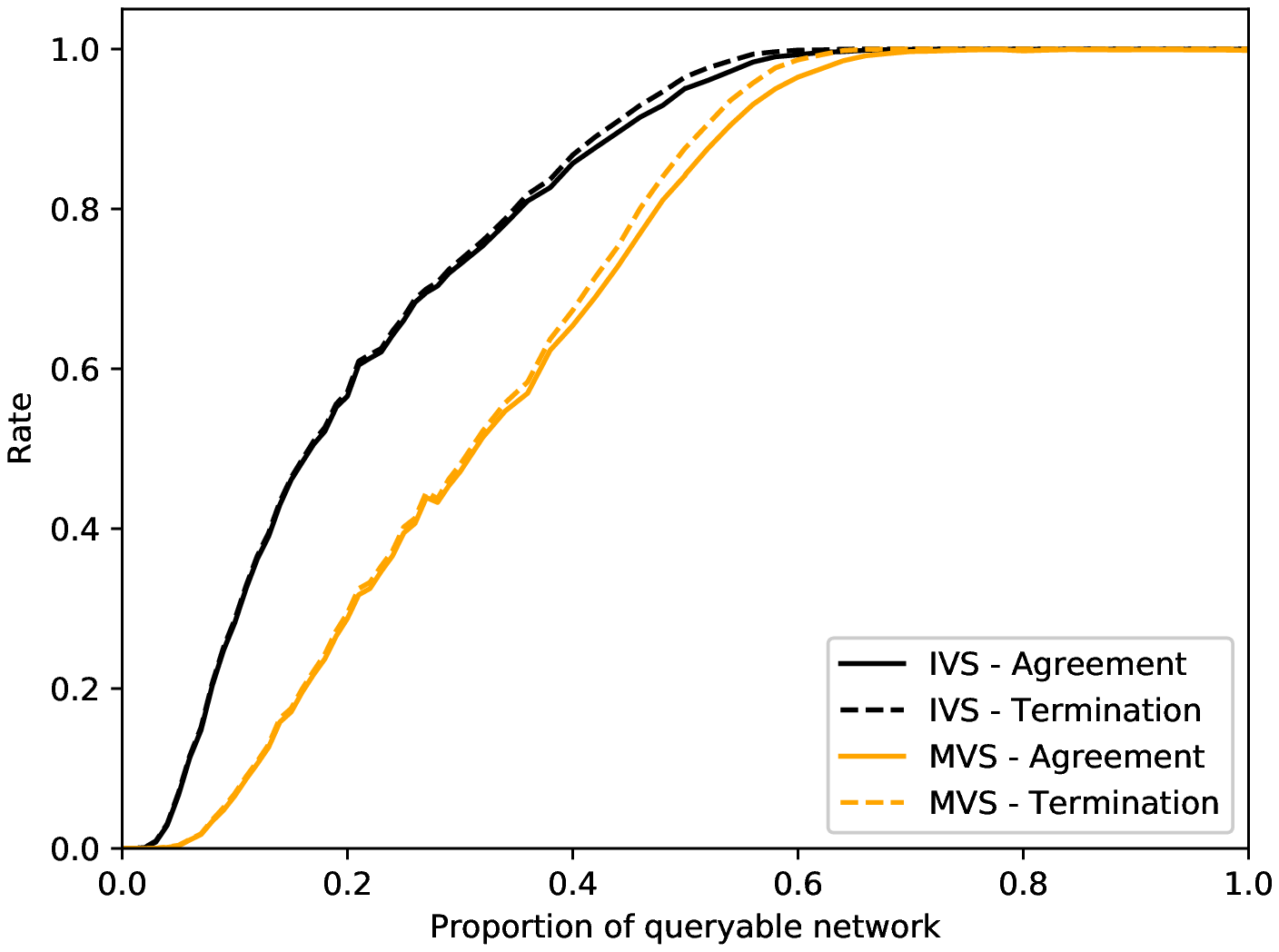}
   \vspace{-0.8cm}
  \captionof{figure}{\small Termination and agreement rate for a Watts-Strogatz graph as a function of the parameter $\delta$.}\label{fig:delta-ATI}
\end{minipage}\hfill
\end{center}

\subsubsection*{Randomization of the threshold}
Figs. \hyperref[fig:q-beta-T]{\ref{fig:q-beta-T}}-\hyperref[fig:q-beta-A]{\ref{fig:q-beta-A}} show the variation of the termination and agreement rate with $q$ and $\beta$. It can be seen that if no randomness is employed the protocol can be prevented from terminating. Introducing randomness via decreasing $\beta$ improves the termination as well as the agreement rate, with an optimum at about $\beta=0.3$. 

\begin{center}
\begin{minipage}{0.45\textwidth}
    \hspace{-1cm}
    \includegraphics[width=1.3\textwidth,trim={0 0 0 0.5cm},clip]{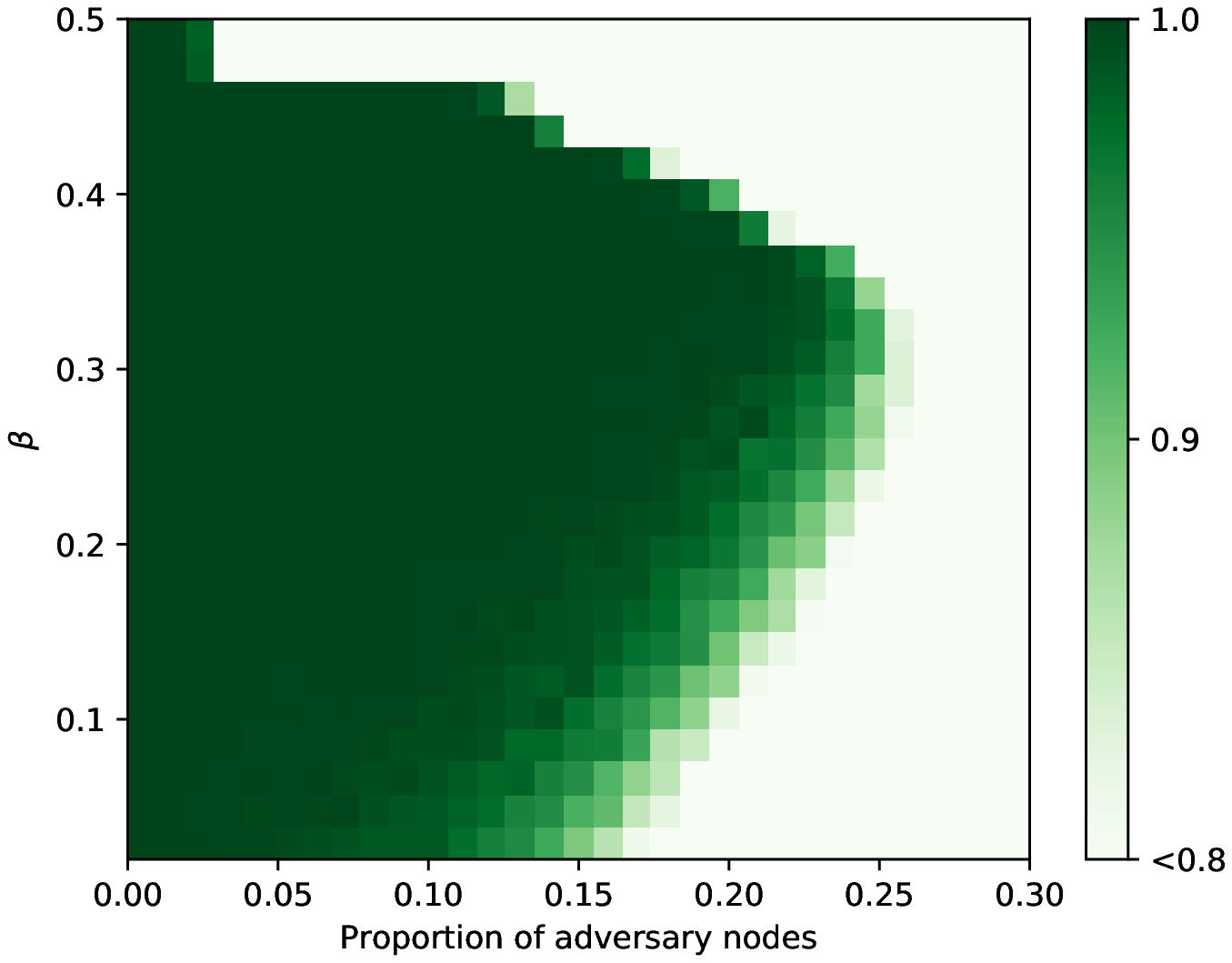}
    \vspace{-0.8cm}
    \captionof{figure}{\small Termination rate}
    \label{fig:q-beta-T}
\end{minipage}\hfill
\begin{minipage}{.45\textwidth}
    \hspace{-1.2cm}
    \includegraphics[width=1.3\textwidth,trim={0 0 0 0.5cm},clip]{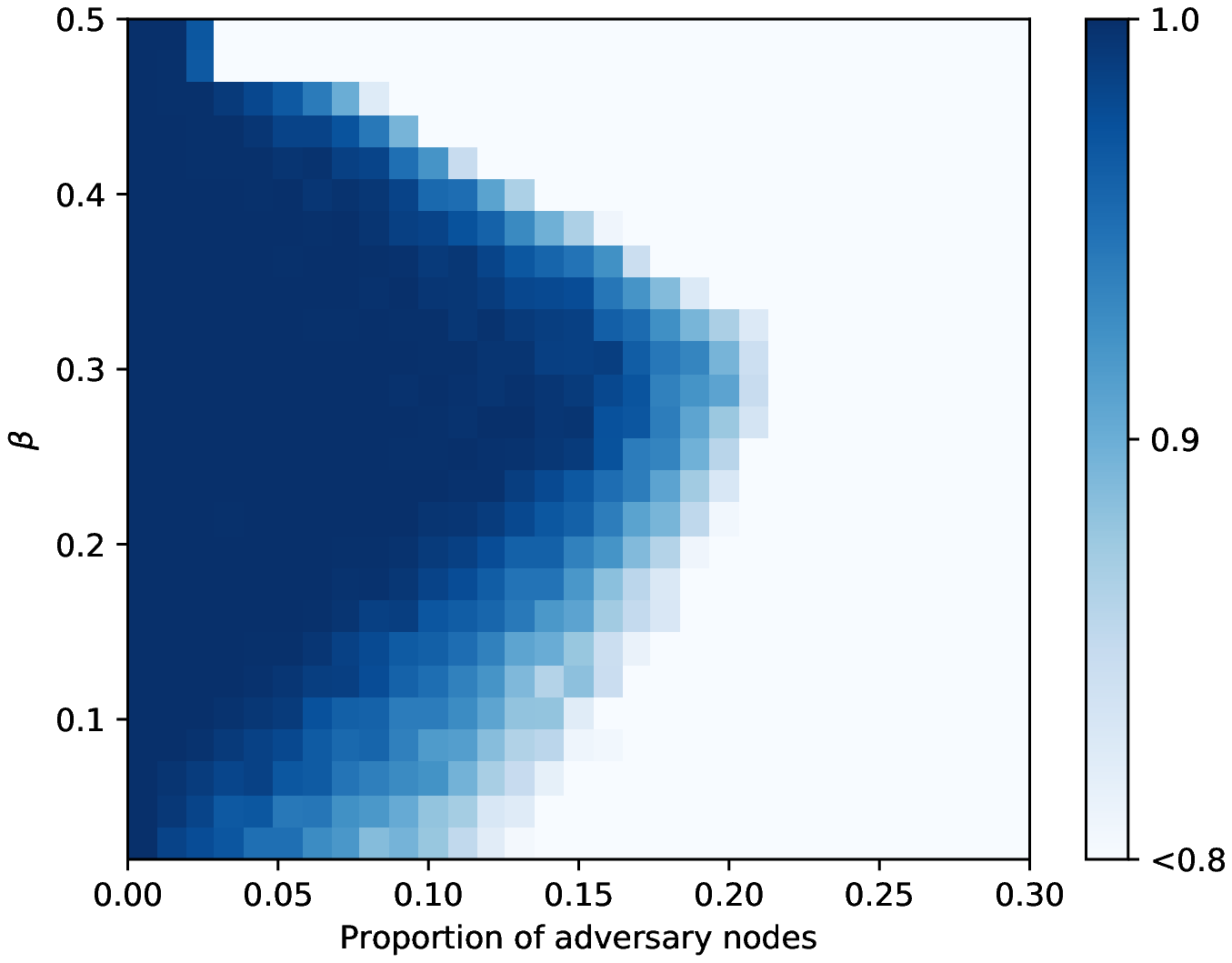}
    \vspace{-0.8cm}
    \captionof{figure}{\small Agreement rate}
    \label{fig:q-beta-A}
\end{minipage}\hfill
\end{center}

We study the dependency of the protocol on the frequency of a random threshold. More precisely, we assume that the protocol uses in a given round a common random number for the threshold with probability $r$, or sets the threshold to 0.5 otherwise. From Fig.\ \hyperref[fig:rand-TAI]{\ref{fig:rand-TAI}}
it can be seen that the agreement and integrity rate increase with $r$, and that in order to achieve a high agreement rate a random number is not required for every round. Furthermore, even just for a very small $r$ the protocol manages to terminate entirely before $\verb?maxIt?$. This also reflects in the total message complexity which is significantly reduced with the introduction of the randomness, as can be seen from Fig.\ \hyperref[fig:rand-rounds]{\ref{fig:rand-rounds}}. 

\begin{center}
\begin{minipage}{0.45\textwidth}
    \hspace{-1cm}
    \includegraphics[width=1.3\textwidth,trim={0 0 0 0.5cm},clip]{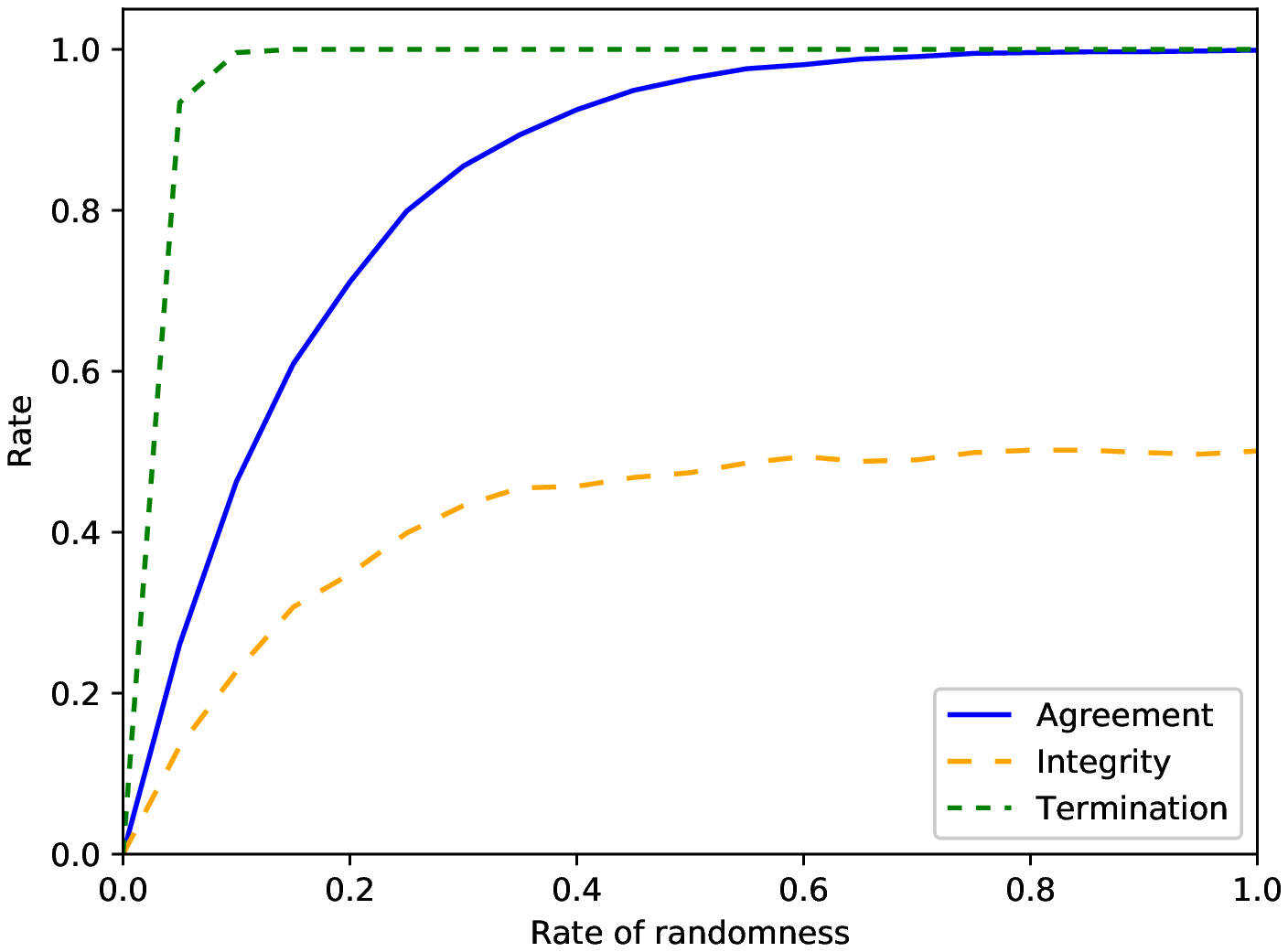}
    \vspace{-0.8cm}
   \captionof{figure}{\small Termination, agreement and integrity rate as a function of the probability $r$ for a round having a random threshold.}
   \label{fig:rand-TAI}
\end{minipage}\hfill
\begin{minipage}{.45\textwidth}
   \hspace{-1.2cm}
    \includegraphics[width=1.3\textwidth,trim={0 0 0 0.5cm},clip]{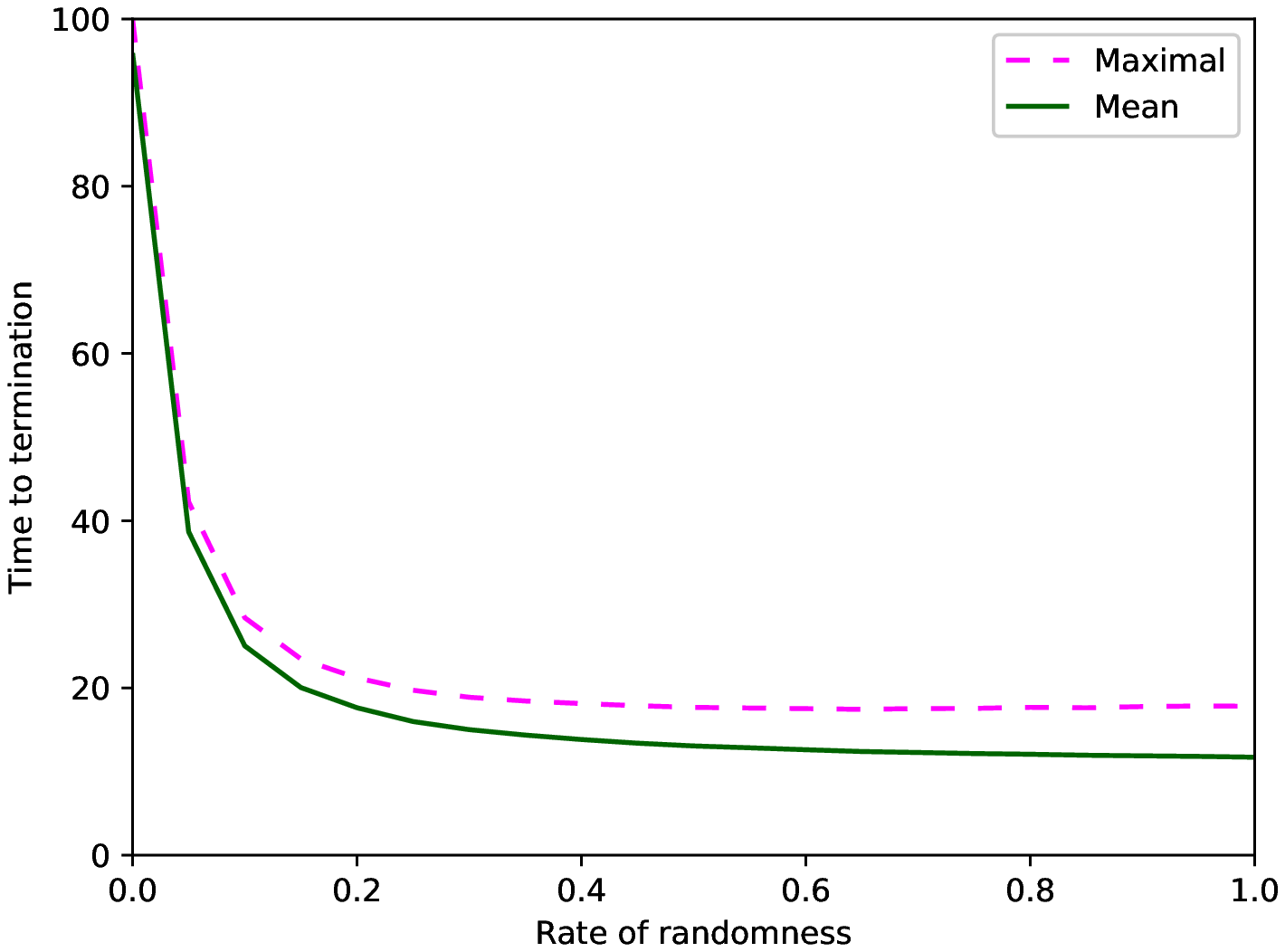}
  \vspace{-0.8cm}
   \captionof{figure}{\small Time to termination as a function of the probability $r$ for a round having a random threshold. \newline}
   \label{fig:rand-rounds}
\end{minipage}\hfill
\end{center}

\subsubsection*{Impact of the initial mean opinion} 
We study how the protocol performs if $p_0$ is different from the first threshold $\tau$. Figs. \hyperref[fig:q-p0-T]{\ref{fig:q-p0-T}}-\hyperref[fig:q-p0-A]{\ref{fig:q-p0-A}} show the termination and agreement rate with $p_0$ and $q$. It can be seen, that as expected the agreement rate is affected most for $p_0$ close to the initial threshold. It can be seen that $p_0$ has no impact on the performance of the protocol below a certain value of $q$. This suggests that the addition of the randomness provides good protection for any mean initial honest opinion. Furthermore and as expected, Fig.\ \hyperref[fig:q-p0-A]{\ref{fig:q-p0-A}} shows that the agreement rate is affected most by this strategy if $p_0$ is close to $\tau$. 

\begin{center}
\begin{minipage}{0.45\textwidth}
   \hspace{-1cm}
  \includegraphics[width=1.3\textwidth,trim={0 0 0 0.5cm},clip]{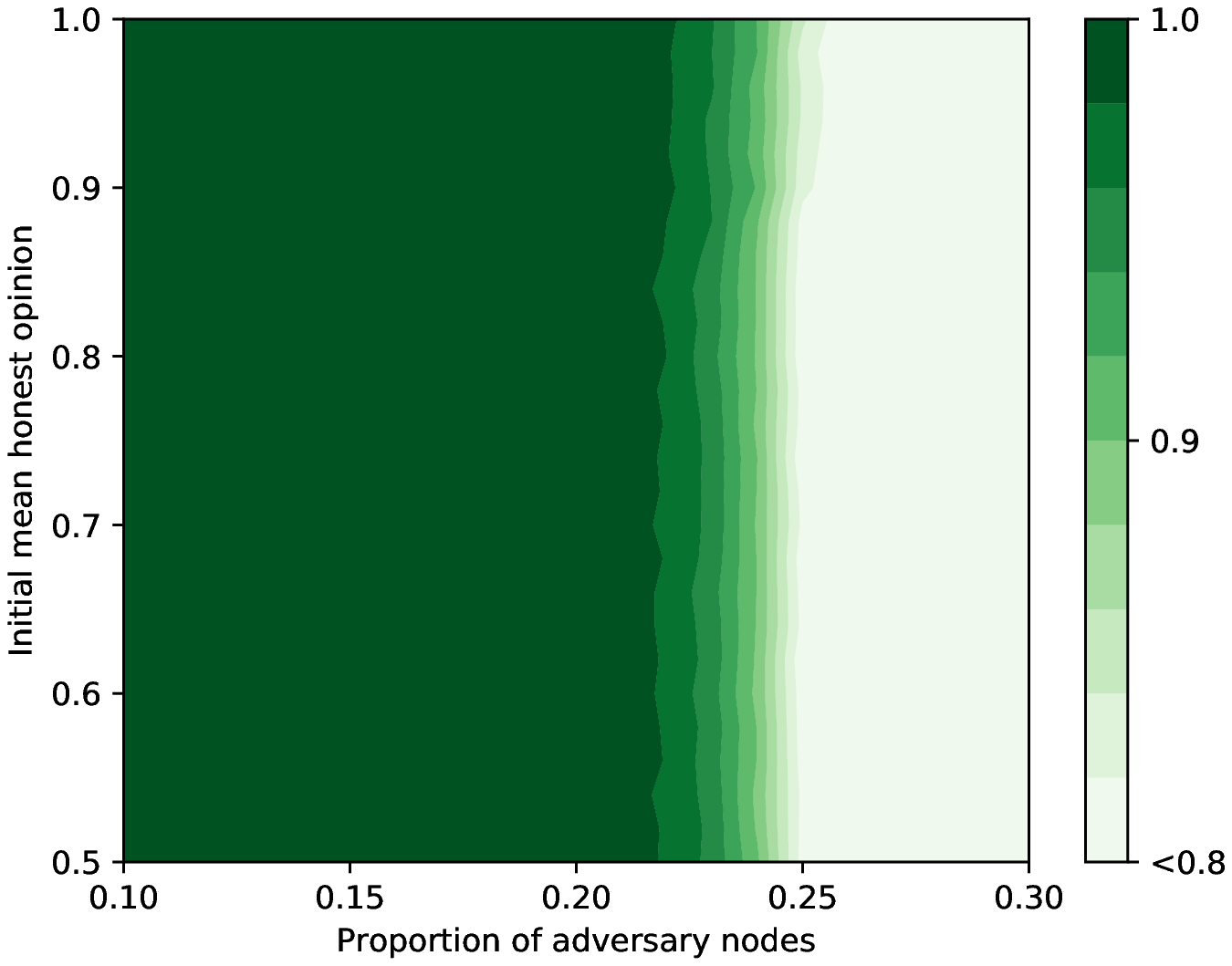}
    \vspace{-0.8cm}
    \captionof{figure}{\small Termination rate}
    \label{fig:q-p0-T}
\end{minipage}\hfill
\begin{minipage}{.45\textwidth}
    \hspace{-1.2cm}
    \includegraphics[width=1.3\textwidth,trim={0 0 0 0.5cm},clip]{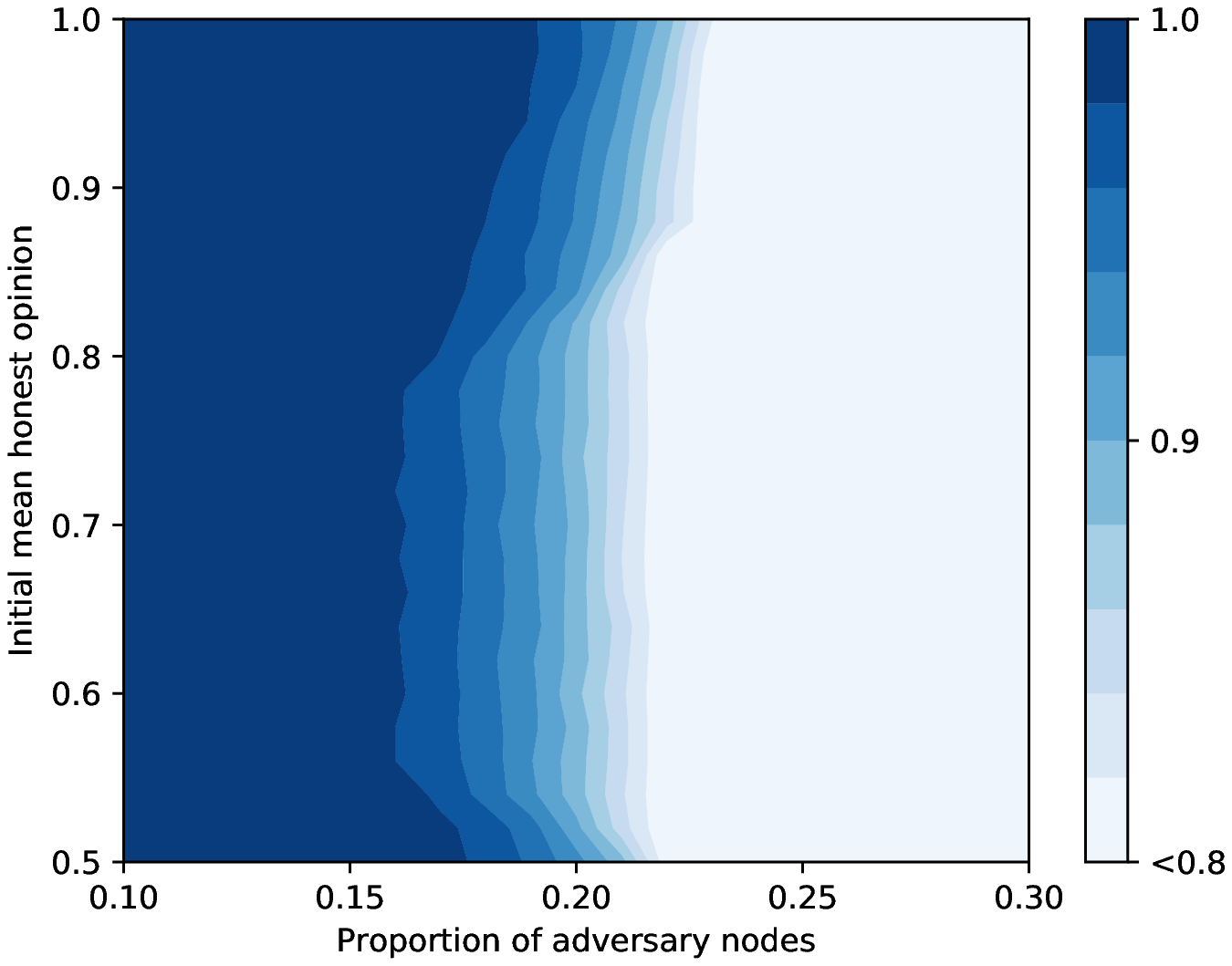}
    \vspace{-0.8cm}
    \captionof{figure}{\small Agreement rate }
    \label{fig:q-p0-A}
\end{minipage}\hfill
\end{center}

\section{Discussion}

\subsection{Synchrony}
In the presence of faults, synchronous and asynchronous systems are fundamentally different  in terms of message or time complexity, but also in terms of what type of problems can be solved. In this paper we focused on the synchronous setting; all nodes update their states simultaneously and the common random number serves as a natural clock signal. Due to the local update rules the majority dynamics and  FPC may also work in an asynchronous setting. However, the FPC will lose the additional feature of the common random threshold. Figures \hyperref[fig:rand-TAI]{\ref{fig:rand-TAI}} and \hyperref[fig:rand-TAI]{\ref{fig:rand-rounds}} indicates that the common random thresholds is only needed at some proportion of the steps.  The nature of the protocol also suggests that the performances of the FPC won't suffer if only some positive fraction of the honest nodes share the same random threshold from time to time. The details on the minimal amount of ``synchronization'' are likely to depend on the precise nature of the system and should be elaborated in additional studies.

\subsection{Generating random numbers}\label{sec:GeneratingRandNum}

The FPC requires the system to generate, from time to time (more precisely, once in each round), a random number available to all the participants (this is very similar to the "global-coin" approach used in many works on Byzantine consensus, see e.g. \cite{AgTo:12}). In the case of distributed systems these random numbers may be provided by a trusted source~\cite{nist:2019}.

We observe that such random number generation can be done in a decentralized way as well (provided that the proportion of the adversarial nodes is not too large) by leveraging on cryptographic primitives such as verifiable secret sharing, threshold signatures or verifiable delay functions, see e.g. \cite{popov-drng:2017, syta:2017, boneh:2018}. 
It is important to observe that, as shown in Figures \hyperref[fig:rand-TAI]{\ref{fig:rand-TAI}} and \hyperref[fig:rand-TAI]{\ref{fig:rand-rounds}}, even if from time to time the adversary can get (total or partial) control of the random number, this can only lead to delayed consensus without agreement or termination failure. Also, it is not necessary that all honest nodes agree on the same number.

\section{Conclusions}\label{sec:conclusions}

In this paper, we compare leaderless binary majority consensus protocols in various faulty and Byzantine infrastructures. We focus on the fast probabilistic consensus protocol (FPC) introduced in \cite{Po:19}. In this protocol, randomness is added to a random majority consensus protocol in order to improve performance in the presence of Byzantine actors. We introduce local stopping rules to enable nodes to conclude individually and automatically. Furthermore, the threshold for majority in the first round is applied at a fixed value $>0.5$ to support asymmetric importance of the 0- and 1-integrity.

We define several adversarial strategies to analyze the performance of the protocol in Byzantine environment. An adversary can be cautious by answering in the same round with the same opinion to everyone, or Berserk by responding to different queries with different opinions. Although the latter may be detectable, we assume in this paper that no such detection mechanism is implemented in the protocol to weaken or prevent a Berserk attack. As a consequence, the Berserk attack can lead to more frequent agreement or termination failures.

We employ simulation tools to study the performance of the protocol with a high statistical significance. In order to assess the performance of the protocol, we study the rate at which the protocol concludes successfully. This is measured with respect to three types of failures: agreement, termination and integrity failure. For the latter, we distinguish also between a 0- and a 1-integrity failure. 

In various real-world situations, nodes may not have a complete network view. We study the protocol when nodes are only capable of querying parts of the network. In order to do so, we employ a Watts-Strogatz model to create a graph that provides a graph with the edges connecting nodes that can query each other. We show that a partial network view of about 50\% is sufficient if we consider the conservative case of a ring graph. We also show that if some of the connections on the ring graph are randomly rewired, the protocol performs well with a much smaller network perception per node. 

From the defined strategies we apply the most efficient one to evaluate the integrity rate. We assume the most conservative case for the 0-integrity, since we analyze this failure by assuming that the initial honest opinion is 49\%. We show that there is a competition between securing the 0- and the 1-integrity rate. Nevertheless, albeit we assume the most conservative case for the 0-integrity the protocol can prevent both integrity failures if the adversary has up to 10\% control over the network. As expected the performance of the protocol improves if the quorum size is increased and it is sufficient to query a fraction of the nodes to avoid an integrity failure. We investigate the performance of the protocol when the network is scaled and it is shown that the protocol performs similarly with the same parameter settings per node. More particularly this means that the average message complexity per node remains constant and that, therefore, the total message complexity in the network increases essentially as $O(n)$. We also show that dependent on the application it may be necessary to reduce the range of the random threshold since the protocol can withstand a higher number of byzantine nodes when the threshold remains close to 0.5. 

We analyze the termination and agreement rate of the protocol for a cautious and Berserk Byzantine strategy and for a worst-case initial setup. We show that the protocol can be prevented from terminating and coming to agreement for an extensive amount of time and that the introduction of the randomness drastically reduces the effects of this attack. We note that for the same proportion of adversarial nodes the Berserk strategy is significantly more severe than the cautious strategy and can reach agreement failure much more efficiently. 
We show that similar to a network occupied by only honest actors the protocol performs well if each node only has a view of the network of about 50\%. Generally, the protocol appears sensitive to the spread of the random thresholds and an optimum parameter set exists. On the other hand, it is also not necessary for the nodes to receive a random number for the random threshold every round, since the performance of the protocol remains the same even if a simple majority rule is applied occasionally.

\bibliography{iota}

\begin{thebibliography}{10}
\bibitem{SaZa:93}
M.~Saks, F.~Zaharoglou, Wait-free k-set agreement is impossible, Proceedings of
  the twenty-fifth annual ACM symposium on Theory of computing - STOC '93.
\end{thebibliography}

\end{document}